%% file: main.tex
\documentclass[journal]{IEEEtran}
\usepackage[utf8]{inputenc}
\usepackage[T1]{fontenc}
\usepackage{url}              
\usepackage{cite}             
\usepackage[cmex10]{amsmath}
\usepackage{amssymb}
\usepackage{xifthen}
\usepackage{xcolor}

\usepackage{mleftright} 
\mleftright  

\usepackage{graphicx}
\usepackage[font=footnotesize]{caption}
\usepackage[font=footnotesize]{subcaption}
\usepackage{tabularx,booktabs}
\newcolumntype{C}{>{\centering\arraybackslash}X} 
\usepackage{algorithmicx}        
\usepackage{algorithm}          
\usepackage{algpseudocode}

\usepackage{stfloats}

\usepackage{amsthm}
\usepackage{mathtools}
\usepackage{multirow}

\input{mathDeclarations}

\begin{document}

\title{On the Computation of the Gaussian Rate-Distortion-Perception Function}

\author{
   \IEEEauthorblockN{Giuseppe Serra,
                    Photios A. Stavrou, \IEEEmembership{Senior Member, IEEE}, and Marios Kountouris, \IEEEmembership{Fellow, IEEE}}
  \thanks{The authors are with the Communication Systems Department at EURECOM, Sophia-Antipolis, France, email: \{\texttt{giuseppe.serra, fotios.stavrou, marios.kountouris\}@eurecom.fr}. This work is part of a project that has received funding from the European Research Council (ERC) under the European Union’s Horizon 2020 Research and Innovation Programme (Grant agreement No. 101003431).}
  }

\maketitle

\begin{abstract} 
In this paper, we study the computation of the rate-distortion-perception function (RDPF) for a multivariate Gaussian source under mean squared error (MSE) distortion and, respectively, Kullback–Leibler divergence, geometric Jensen-Shannon divergence, squared Hellinger distance, and squared Wasserstein-2 distance perception metrics. To this end, we first characterize the analytical bounds of the scalar Gaussian RDPF for the aforementioned divergence functions, also providing the RDPF-achieving forward ``test-channel'' realization. Focusing on the multivariate case, we establish that, for tensorizable distortion and perception metrics, the optimal solution resides on the vector space spanned by the eigenvector of the source covariance matrix. Consequently, the multivariate optimization problem can be expressed as a function of the scalar Gaussian RDPFs of the source marginals, constrained by global distortion and perception levels. Leveraging this characterization, we design an alternating minimization scheme based on the block nonlinear Gauss–Seidel method, which optimally solves the problem while identifying the Gaussian RDPF-achieving realization. Furthermore, the associated algorithmic embodiment is provided, as well as the convergence and the rate of convergence characterization. 
Lastly, for the ``perfect realism'' regime, the analytical solution for the multivariate Gaussian RDPF is obtained. We corroborate our results with numerical simulations and draw connections to existing results.
\end{abstract}

\section{Introduction}
The foundational principles of rate-distortion theory, established by Shannon in \cite{shannon:59}, introduce the idea of a trade-off between the desired bit rate used for encoding or compressing source messages and the resulting achievable distortion between the source message and its reconstructed representation. This highly relevant problem is shown to have a mathematical representation given by the rate-distortion function (RDF) and has set the cornerstone for the development of lossy compression algorithms across various multimedia applications. Nevertheless, in recent years, a body of research spanning from machine learning and computer vision to multimedia applications, see e.g.,  \cite{theis:2016:hyperRes, shaham:2018:defImCompr, kudo:2019, minnen:2018, chen:2021:improc_ml}, has highlighted the limitations of solely focusing on distortion minimization in the reconstructed signals. 
Empirical evaluations of reconstruction quality by means of human scoring \cite{mentzer:2020, moorthy:2011} show that conventional distortion measures fail to capture human preference and perception, especially in extreme compression scenarios. Therefore, perceptual quality, which refers to the property of a sample to appear pleasing from a human perspective, cannot be guaranteed by the conventional distortion minimization alone.

Recognizing the need for a comprehensive characterization of the RDF that encompasses perceptual quality, Blau and Michaeli in \cite{blau:2019} introduced a generalization of the RDF, which they called rate-distortion-perception function (RDPF). Around the same time, a similar problem appeared in two companion papers by Matsumoto in \cite{matsumoto:2018,matsumoto:2019}. The RDPF extends the classical single-letter RDF formulation by incorporating a divergence constraint between the source distribution and its estimation at the destination. This divergence constraint serves as a proxy for human perception, measuring the degree of satisfaction experienced when employing the data. 
The same idea of quantifying the deviation from what is refereed to as {\it natural scene statistics} has previously been employed in numerous no-reference image quality metrics, showing strong correlations with human opinion scores \cite{mittal:2013, saad:2012}. Furthermore, this principle is the underlying mechanism behind generative adversarial network (GAN)-based image compression and restoration models, where heightened perceptual quality is achieved by directly minimizing a certain divergence measure between source samples and their reconstructions \cite{mentzer:2020, agustsson:2019, wang:2018esrgan}. However, it is worth noting that identifying the divergence function that most effectively aligns with human perception remains a field of active research.

On the other hand, as initially hinted in \cite{kountouris:2020}, a divergence measure can also be viewed as a semantic quality metric, reflecting the relevance of the reconstructed source from the observer's perspective. 
For example, Katakol {\it et. al.} in \cite{katakol:2021} compare the performance in the segmentation task of models trained on traditionally compressed samples against compressed samples whose perceptual quality has been enhanced using GAN-based restoration models. The results show gains in the task quality metric, especially in the segmentation scores of smaller scene objects, usually more susceptible to the introduced distortion.
Another relevant yet different setup has recently been introduced in \cite{stavrou:2023}, in which in place of the perception quality, there
exists an additional distortion criterion instead of a divergence.

\subsection{Related Work}\label{subsec:literature_review}
\par In the realm of the rate-distortion-perception framework, Theis and Wagner in \cite{theis:2021} provide a coding theorem for stochastic variable-length codes in both one-shot and asymptotic regimes, assuming infinite common randomness between the encoder and decoder of the RDP problem, and exploiting properties of the strong functional representation lemma \cite{li:2018}. Chen {\it et. al.} in \cite{chen:2022} derive coding theorems for the asymptotic regime, analyzing the operational meaning of the RDPF for three distinct cases; when the encoder and the decoder share or not common randomness, and when both have private randomness. Li \textit{et. al.} in \cite{li:2011:distribution} study the special case of the \textit{``perfect realism''}, that is to say, the extreme situation where the source distribution is constrained to be the same as the reconstruction distribution, in the context of distribution preserving quantization and distribution preserving RDF. They derived an achievability result for the Gaussian case. Recently, Wagner in \cite{wagner:2022:rate} provides a coding theorem for the RDPF trade-offs for the perfect and near perfect realism cases, when only finite common randomness between encoder and decoder is available. 

Similar to the classical RDF, the RDPF defined for general sources does not enjoy any analytical solution. Nevertheless, there exist some closed-form expressions for specific classes of sources, such as binary sources subject to Hamming distortion and total variation distance \cite{blau:2019} and scalar Gaussian sources under mean squared-error (MSE) distortion and squared Wasserstein-2 distance \cite{zhang:2021}. In \cite{qian:2023} the authors have recently derived closed-form parametric expressions by means of a reverse water-filling algorithm for the case of the multivariate Gaussian source under MSE distortion when the perception constraint is either the squared Wasserstein-2 distance or the Kullback–Leibler (KL) divergence.

The difficulty in deriving analytical solutions for RDPF have stimulated the study of computational methods for its estimation. Toward this end, Serra {\it  et al.} in \cite{serra:2023} characterize a generalization of the celebrated Blahut-Arimoto algorithm \cite{blahut:1987} to compute the RDPF for general discrete sources under a generic single-letter distortion metric, and a perception constraint that belongs to the family of $f$-divergences (details on this rich class of divergences can be found for instance in \cite{sason:2018}). Always for discrete sources, Chen {\it et. al.} in \cite{chen:2023} reformulate the RDPF problem  as a Wasserstein barycenter problem for specific cases of Wasserstein distances, KL divergence, and total variation distance and provide a computational method using a variation of the Sinkhorn algorithm. Alternatives to these algorithmic approaches for the computation of the RDPF rely on data driven solutions, which employ generative adversarial networks minimizing a linear combination of distortion and perception metrics, see e.g., \cite{blau:2019, zhang:2021, ogun:2021}. Nevertheless, despite providing a practical framework for data driven codec optimization, these methodologies are highly computational and data intensive, while lacking generalization capabilities. 

\subsection{Our Approach and Contributions}\label{subsec:contributions}
The objective of this work is twofold. First, we aim to derive analytical bounds (for some cases we prove that these are exact) of the RDPF under MSE distortion when the perception constraint belongs to certain well-known and widely-used divergence, that is, the KL divergence \cite{sason:2018}, the geometric Jensen-Shannon divergence \cite{nielsen:2019}, the squared Hellinger distance \cite{sason:2018}, and the squared Wasserstein-2 distance \cite{gelbrich:1990:W2formula} for scalar-valued Gaussian random variables. The second and most important contribution is the construction of a generic algorithmic approach for the optimal computation of the multivariate Gaussian RDPF under MSE distortion constraint and any member of the aforementioned class of perception constraints (or any other divergence that admits closed-form expression for Gaussian random variables). To summarize, in this paper we derive the following new results:
\begin{itemize}
    \item In Section \ref{sec: unvariate closed forms}, we characterize closed-forms expressions of the scalar Gaussian RDPF for direct or reverse KL divergence, the geometric Jensen-Shannon divergence, the squared Hellinger distance, and the squared Wasserstein-2 distance perception constraints. 
    \item In Section \ref{subsec:generic_algo}, we prove that, under tensorizable distortion and divergence constraints, the optimal solution of the multivariate Gaussian RDPF subject to an MSE distortion and {\it any perception constraint that is convex on the second argument}, can be found on the space of the eigenvectors of the source covariance matrix. In other words, the problem of convex Gaussian RDPF under MSE distortion and any divergence perception constraint achieves an optimal solution when the involved covariance matrices commute by pairs \cite[Section 0.7.7]{horn:2012:matrix}. The resulting optimization problem can be solved optimally using an alternating minimization approach, by means of the {\it block nonlinear Gauss–Seidel method} \cite{grippo:2000}, for which we also develop its algorithmic embodiment (see Algorithm \ref{alg: AM}). For the specific algorithm, we show convergence (Theorem \ref{th:gmrdpf:RDPF_AM}) and provide an upper bound on the worst-case convergence rate (Theorem \ref{th:conv:convrate}). 
    \item In Section \ref{subsec:applicationAlg}, we provide as an application example, the implementation of Algorithm \ref{alg: AM} for the MSE distortion and squared Wasserstein-2 divergence constraints. Although only the solution of one of the subproblems of Algorithm \ref{alg: AM} is available in closed form (Theorems \ref{th:gmrdpf:AM_subD}), the complementary subproblem can be solved optimally through numerical methods (Theorem \ref{th:gmrdpf:AM_subP}). We note that similar implementations can be derived for the other divergence constraints that are used in this paper via the results summarized in Table \ref{table:subDClosedForm}.
    \item In Section \ref{subsec: perfect_realism}, we leverage the analytical results obtained from the alternating minimization approach that led to Algorithm \ref{alg: AM}, to characterize the closed-form solution of the multivariate Gaussian RDPF in the regime of \emph{perfect realism} (Corollary \ref{th: perfect_realism}). Specifically, this result provides the optimal stagewise distortion allocation, i.e., the distortion introduced on each dimension of the Gaussian reconstruction, according to which can be interpreted as an {\it adaptive water-level} (see Fig. \ref{fig: comparison_per_dimension}). In fact, unlike the reverse water-filling solution in the classical RDF problem, the obtained solution presents dependency on the stagewise second order moment of the source. 
\end{itemize}   

We highlight that for the KL divergence and the squared Wasserstein-2 distance in the scalar case, we solve the problem differently from \cite{zhang:2021, qian:2023}  in a way that allows to also find the optimal linear encoder and decoder that achieve the specific closed-form expression (i.e., the optimal forward test-channel realization). When it comes to the vector Gaussian case, we note that the reverse water-filling parametric solutions in \cite{qian:2023} are only applicable to specific problems. In contrast, our alternating minimization approach and the corresponding algorithmic embodiment obtained here can be applied to any Gaussian RDPF problem with MSE distortion, as long as the scalar Gaussian RDPF admits a characterization. In addition, our approach provides, similar to the scalar case, the optimal linear encoder and decoder pair that achieves the Gaussian RDPF as a result of the optimization procedure.

\subsection{Notation}
Given a Polish space $\mathcal{X}$, we denote by $(\mathcal{X}, \mathbb{B}(\mathcal{X}))$ the Borel measurable space induced by the metric, with $\mathcal{P}(\mathcal{X})$ denoting the set of probability measures defined thereon. 
Given $p, q \in \mathcal{P}(\mathcal{X})$, we denote with $p \ll q$ the absolute continuity of $p$ with respect to $q$, meaning that given a set $\mathcal{A} \in \mathbb{B}(\mathcal{X}), q(\mathcal{A}) = 0 \implies p(\mathcal{A}) = 0$.
For a random variable $X$ defined on $(\mathcal{X}, \mathbb{B}(\mathcal{X}))$, we denote with $\pdf{X} \in \mathcal{P}(\mathcal{X})$ its probability measure and with $\mean[X]$ and $\Cov[X]$ its mean and covariance matrix, respectively. 
\par The identity matrix on $\mathbb{R}^{N \times N}$ is denoted by $I$. Given a matrix $A \in \mathbb{R}^{N \times N}$, we will indicate with $s_A = [s_{A,i}]_{i=1:N}$ the vector of its singular values and, if $A$ is diagonalizable,  with $\lambda_A = [\lambda_{A,i}]_{i=1:N}$ the vector of its eigenvalues. However, in the cases where the notation $s_A$ or $\lambda_A$ is inadequate, we indicate the eigenvalues and singular values as operators $\lambda[A]$ {and} $\sv[A]${, respectively}. 

Furthermore, we introduce the notations $\lambda^{\downarrow}[\cdot]$ {or}  $\sv^{\downarrow}[\cdot]$ and $\lambda^{\uparrow}[\cdot]$ {or} $\sv^{\uparrow}[\cdot]$, {respectively,} for the vectors whose coordinates are the {eigenvalues or singular values} arranged in {a} decreasing order {or in an} increasing order. We indicate positive definiteness (resp. positive semi-definiteness) with the notation $A \succ 0$ (resp. $A \succeq 0$). The trace operator will be denoted by $\Tr[\cdot]$, while $|| \cdot ||_1$, $||\cdot||_F$, and $\langle\cdot,\cdot\rangle_F$ indicate, respectively, Schatten-1 norm \cite[Section 4.2]{bhatia:2013}, Frobenius norm and Frobenius inner product \cite[Equation 5.2.7]{horn:2012:matrix}. A closed interval on the set of integers $\mathbb{N}$ is indicated by $a:b$, where $a$, $b$ are the endpoints of the set. For any set $S \subseteq \mathbb{R}^n$, we define its complement set by $\mathcal{S}^c \triangleq \{ x \in \mathbb{R}^n: x \notin \mathcal{S} \}$. The identity function is denoted as $\id(\cdot)$. The Lambert W function \cite{corless:1996:lambert} is denoted by $W(\cdot)$, whereas its primary and secondary branch are denoted as $\lamb[0](\cdot)$ and $\lamb[-1](\cdot)$, respectively.
Given a real function $(x,y) \xrightarrow{} f(x,y)$ with $x \in \mathbb{R}^n$, $y \in \mathbb{R}^m$, we indicate with $\nabla f$ the gradient of the function $f(x,y)$ and with $\nabla_x f$ (or $\nabla_y f$) the gradient with respect to the specific variable.\\

\section{Preliminaries on RDPF and Certain Divergences}\label{sec:preliminaries}

{In this section, we first give some preliminary results on RDPF, following the works of \cite{blau:2019,theis:2021,chen:2022}. Subsequently, we discuss the divergence metrics used in the paper. }

{\subsection{RPDF}\label{subsec:rdpf}}

{We start with the definition of the RDPF defined for general alphabets.}

\begin{definition}(RDPF) \label{def:rdpf}
Let a source $X$ be a random variable on $(\mathcal{X}, \mathbb{B}(\mathcal{X}))$ distributed according to the probability measure $\pdf{X} \in \mathcal{P}(\mathcal{X})$.
Then, the RDPF for a source {$X\sim{p}_X$} under {the} distortion measure $\Delta:\mathcal{X}^2 \to \mathbb{R}^+_0$ and divergence function $d: \mathcal{P}(\mathcal{X}) \times \mathcal{P}(\mathcal{X}) \to  \mathbb{R}^+_0$ is defined as {follows:}
\begin{align}
        R(D,P) = & \min_{\cpdf{\hX}{X}}  I(X,\hX) \label{opt:RDPF} \\
        \textrm{s.t.}   & \quad \E{\Delta(X,\hX)} \le D \label{opt: distortion_constraint}\\
                        & \quad d(p_X||p_{\hX}) \le P \label{opt: perception_constraint} 
\end{align}
where the minimization is among all conditional distributions ${\cpdf{\hX}{X}}: \mathcal{X} \to \mathcal{P}(\mathcal{\hat{X}})$.

\end{definition}
The RDPF generalizes the rate-distortion (RD) function, complementing the inherited single letter distortion constraint \eqref{opt: distortion_constraint} with a divergence constraint \eqref{opt: perception_constraint} between the source distribution $\pdf{X}$ and the reconstruction induced distribution $\pdf{\hat{X}}$.

\begin{remark}\label{remark:1}(On Definition \ref{def:rdpf}) Following \cite{blau:2019}, it can be shown that \eqref{opt:RDPF} has some useful {functional} properties, under mild regularity conditions.
In particular, \cite[Theorem 1]{blau:2019} shows that $R(D,P)$ is (i) monotonically non-increasing function in both $D\in[D_{\min},D_{\max}]\subset[0,\infty)$ and $P\in[P_{\min},P_{\max}]\subset[0,\infty)$; (ii) convex if the divergence $d(\cdot||\cdot)$ is convex in its second argument. 
\end{remark}

In the following, we provide known results on the operational meaning of the RDPF, linking the RDPF to the fundamental compression limit of a source under suitable distortion and perception constraints. Throughout the paper, we assume that we are given an $\IID$ sequence of $N$-length random variables ${X}^N$ with $ X_i \sim \pdf{X} \in \mathcal{P}(\mathcal{X})$, {and proceed to define the notions of \texttt\{{\it encoder, decoder, code}\} functions following \cite{theis:2021}}. 

\begin{definition}
    For an arbitrary set $\mathcal{X}$, a (possibly stochastic) encoder is defined as any function belonging to  $\mathcal{F} = \{f: \mathcal{X}^N \times \mathbb{R} \to \mathcal{M}\}$. Similarly, a (possibly stochastic) decoder is a function in $\mathcal{G} = \{ g: \mathcal{M} \times \mathbb{R} \to \mathcal{X}^N\}$. A code is defined as an element of $\mathcal{F} \times \mathcal{G}$.
\end{definition}

We can now introduce the definition of achievability and that of {the} infimum of all achievable rates.

\begin{definition}(Achievability) \label{def: RDPF Achievability}
Given a distortion level $D\geq{0}$ and a perception constraint $P\geq{0}$, a rate $R$ is said to be $(D,P)$-achievable if, for sufficiently large $N$, there exist a common random variable $U$ on $\mathbb{R}$ and a stochastic lossy source code in $\mathcal{F} \times \mathcal{G}$ such that:
\begin{align}
M = f(X^N, U) \qquad \hat{X}^N = g(M, U) \nonumber
\end{align}
satisfying:
\begin{align}
    &\frac{1}{N}H(M|U) \le R \nonumber \\ 
    &\frac{1}{N} \sum_{i=1}^N \E{\Delta(X_i,\hat{X}_i)} \le D \nonumber \\
    &d( \pdf{X_i} || \pdf{\hat{X}_i}) \le P \qquad i \in 1:N. \label{constr: per_letter_perception}
\end{align}
Furthermore, assuming the previous conditions to be satisfied, we define {the operational rates as follows:}
\begin{align}
    R_{cr}(D,P) \equiv \inf\{R: (R,D,P) \text{ is achievable}\}.\label{eq: rate_inferior_cr} \nonumber
\end{align}
\end{definition}

Theis and Mikaeli in \cite[Theorem 3]{theis:2021}, {provided} a coding theorem {that links} the operational definition of $R_{cr}(D,P)$ with the {information definition of} $R(D,P)$. {This is stated next.}
\begin{theorem} \label{th: rdp_achivable_asymptotic}
For $D\geq{0}$, $P \ge 0$, we obtain
    \begin{align}
        R_{cr}(D,P)=R(D,P).
    \end{align}    
\end{theorem}

{Recently, Chen {\it et al.} in \cite[Remark 3]{chen:2022} stressed that provided Remark \ref{remark:1}, (ii) holds for the definition of RDPF, the results of Theorem \ref{th: rdp_achivable_asymptotic} remain valid even if 
the per-letter perception constraints \eqref{constr: per_letter_perception} are weakened to
    \begin{align*}
        \frac{1}{N} \sum_{i=1}^N d(\pdf{X_i}||\pdf{\hat{X}_i})
        \quad \text{or} \quad
        d \left(\pdf{X}\Biggl{|}\Biggl{|} \frac{1}{N} \sum_{i=1}^N \pdf{\hat{X}_i} \right).
    \end{align*}
}

\subsection{Preliminaries on Certain Classes of Divergence Functions}
{In what follows, we provide an introduction to the divergences that will be used throughout the paper.}

{\textbf{Kullback–Leibler (KL) divergence $(\KL)$.} The KL divergence is defined as follows:}
\begin{align*}
    \KL(\pdf{X}|| \pdf{\hX}) {\triangleq} \int_{\mathcal{X}} \pdf{X}(u) \log{\frac{\pdf{X}(u)}{\pdf{\hX}(u)}} du \quad \pdf{X},\pdf{\hX} \in \mathcal{P}(\mathcal{X})
\end{align*}
where $\KL(\pdf{X}|| \pdf{\hX})$ is finite if and only if $\pdf{X} \ll \pdf{\hX}$. It should be remarked that, in general,  $\KL(\pdf{X}||\pdf{Y}) \neq \KL(\pdf{Y}||\pdf{X})$ \cite[Remark 20]{sason:2018}. For this reason, in the sequel, we distinguish between the two cases by addressing as {\it direct} the function $\KL(\pdf{X}|| \cdot)$ and as {\it reverse} the function $\KL(\cdot||\pdf{X})$, depending on the placement of the source distribution $\pdf{X}$. 

\rem{\subsubsection{Geometric Jensen-Shannon (GJS) divergence $\GJS$}}

\textbf{Geometric Jensen-Shannon (GJS) divergence $(\GJS)$.}
Following \cite[Theorem 2]{nielsen:2019}, next we give the definition of GJS divergence.
\begin{definition}[GJS Divergence] \label{def: GJS divergence}
Given the induced distributions {of} the random variables $X \sim p_X$ and ${\hX} \sim p_{\hX}$, the {GJS} is defined as follows:
\begin{align}
    \GJS(p_X||p_{\hX}){\triangleq} \frac{1}{2} \KL(p_X||p_g) + \frac{1}{2} \KL(p_{\hX}||p_g) \label{eq:GJS_general}
\end{align}
where $p_g$ is the geometric mean between $p_X$ and $p_{\hX}$ defined as $p_g = \frac{1}{Z}(p_X)^\frac{1}{2}(p_{\hX})^\frac{1}{2}$, with $Z = \int_{-\infty}^{+\infty}du (p_X(u))^\frac{1}{2}(p_{\hX}(u))^\frac{1}{2}$. If $p_X \sim \ND(\mean[X],\Sigma_X)$ and $p_{\hX} \sim \ND(\mean[\hX],\Sigma_{\hX})$, then $p_g$ is a Gaussian distribution whose mean $\mean[g]$ and covariance $\Sigma_g$ are given by
\begin{align}
    \Sigma_g = \left( \frac{1}{2} \Sigma_X^{-1} + \frac{1}{2} \Sigma_{\hX}^{-1} \right)^{-1}\\
    \mean[g] = \Sigma_g \left(\frac{1}{2} \Sigma_X^{-1}\mean[X] + \frac{1}{2} \Sigma_{\hX}^{-1}\mean[\hX] \right).
\end{align}
\end{definition}
{It should be noted that the GJS divergence in Definition \ref{def: GJS divergence} is a close relative of the classical Jensen-Shannon (JS) divergence (see, e.g., \cite{sason:2018}), and deviates from the latter in that the symmetrization of the KL divergence is induced by a geometric mean instead of an arithmetic mean. This difference allows the GJS divergence to admit a closed-form expression for Gaussian distributions, as opposed to the classical JS divergence.}

\textbf{Squared Hellinger distance $\HS$.} Following \cite{sason:2018}, the $\HS$ distance is defined as:
\begin{align}
    \HS(\pdf{X},\pdf{\hX}) {\triangleq} \frac{1}{2} \int_{\mathcal{X}} \left( \sqrt{\pdf{X}(u)} - \sqrt{\pdf{\hX}(u)} \right)^2 du, 
\end{align} \label{eq:HS_general}
i.e., the Hellinger distance is the $\ell^2$ norm between $\sqrt{\pdf{X}}$ and $\sqrt{\pdf{\hX}}$, therefore respecting the following upper-bound
\begin{align*}
    \HS(\pdf{X},\pdf{\hX}) \le ||\sqrt{\pdf{X}}||^2_2 + ||\sqrt{\pdf{\hX}}||^2_2 = 2.
\end{align*}

\textbf{Squared Wasserstein-2 distance $\WTS$.}  {Squared Wasserstein distance was originally introduced in \cite{gelbrich:1990:W2formula} and is known to have strong connections to the optimal transport problem (see, e.g., \cite[Chapter 7]{villani:2021})}. {In particular, squared Wasserstein-2 distance  is defined as follows:
\begin{align}
    \WTS(\pdf{X},\pdf{\hX}) {\triangleq} \min_{\Pi} \E{||X - {\hX}||^2}
\end{align}
where $\Pi$ is the set of all joint distributions {$p_{(X,\hX)}$} with {given} marginals $\pdf{X}$ and $\pdf{\hX}$.}

{\section{RDPF for scalar-valued Gaussian sources}\label{sec: unvariate closed forms}}

{In this section, we derive closed-form expressions of the RDPF for a scalar-valued Gaussian source under MSE distortion constraint and various perception constraints. The considered class of perception constraints consists of direct or reverse $\KL$, $\GJS$, $\HS$, and $\WTS$ perception constraints\footnote{For the squared Wasserstein-2 divergence, the same closed-form solution is derived using a different methodology in \cite[Theorem 1]{zhang:2021}.}, respectively. Moreover, on top of the closed-form expressions of Gaussian RDPF for the previous class of perception constraints, our methodology completely specifies the design variables of the {\it optimal linear realization} (i.e., the optimal selection of encoder and decoder policies) that achieve the obtained Gaussian RDPF bounds for each perception constraint.}

In what follows, we characterize \eqref{def:rdpf} for jointly Gaussian random variables and MSE distortion constraint.
\begin{problem}\label{problem:1}
Given a Gaussian source $X \sim \ND(\mean[X], \var[X])$, $\var[X]>0$, assume that the reconstructed message $\hat{X}$ is chosen such that the joint tuple $(X,\hat{X})$ is also Gaussian. Then, the reconstructed message admits a linear (forward) realization of the form $\hat{X} = aX + W$, where $a\in\mathbb{R}$, $W \sim \ND(\mean[W], \var[W])$, $\var[W]\ge{0}$, such that $\mean[\hX] = a\mean[X] + \mean[W]$ and $\var[\hX] = a^2\var[X] + \var[W]$. By considering the case where $\mean[X]=\mean[\hX]$ (by setting $\mean[W] = (1 - a)\mean[X]$), and $\E{\Delta(X,\hX)}\equiv\E{(X-\hX)^2}$, we can cast \eqref{opt:RDPF}-\eqref{opt: perception_constraint} as follows:
\begin{align}
    \begin{split}
        R(D,P) &=  \min_{a\in\mathbb{R},\var[W]\ge{0}} \tfrac{1}{2} \log \left(1 + a^2 \frac{\var[X]}{\var[W]} \right)\\
        \textrm{s.t.}   & \quad (1-a)^2\var[X] + \var[W] \le D \\
                    & \quad d(\pdf{X},\pdf{\hX}) \le P
    \end{split} \label{opt: uRDPGaussian}
\end{align}
where the specific form of $d(\cdot,\cdot)$ depends on the choice of the perception constraint.
\end{problem}
We point out the following technical remark on Problem \ref{problem:1}. 
\begin{remark}(On Problem \ref{problem:1}) If in Problem \ref{problem:1} the choice of the perception constraint is restricted to $\WTS(\cdot,\cdot)$ or the reverse $\KL(\cdot||\cdot)$, it can be shown that their respective minimizing distribution is itself Gaussian (see e.g., \cite{fang:2020} and \cite[Appendix A]{zhang:2021}). This guarantees that \eqref{opt: uRDPGaussian} in these two cases provides an exact characterization of the Gaussian RDPF. Alas, the same property does not hold, in general, for many divergences such as the cases of direct-$\KL(\cdot||\cdot)$, $\GJS(\cdot||\cdot)$ or $\HS(\cdot||\cdot)$. In such cases, the characterization in \eqref{opt: uRDPGaussian} serves as an upper bound to the optimal solution.
\end{remark}
\par {In the following theorem, we compute in closed-form the characterization of the Gaussian RDPF defined in Problem \ref{problem:1}, when $d(\cdot,\cdot)$ corresponds to the direct $\KL(\pdf{X}||\pdf{\hX})$.}

\begin{theorem}[{RDPF under direct $\KL(\cdot||\cdot)$}] \label{th:uRDP:KL}
{Consider the characterization in Problem \ref{problem:1} with $d(\cdot,\cdot)\equiv{D}_{KL}(\pdf{X}||\pdf{\hX})$. Then the solution of \eqref{opt: uRDPGaussian} is obtained analytically, as follows: }
\begin{align} 
    \begin{split}
    &R(D,P) = \\ 
    &   \begin{cases}
            \max \left\{ \tfrac{1}{2} \log \left( \frac{\var[X]}{D} \right), 0 \right\} \quad \text{if } (D,P) \in \mathcal{S}^c\\
            \tfrac{1}{2} \log \left(1 + \tfrac{\var[X] \left( 1 - \tfrac{D}{\var[X]} - g(P) \right)^2}{4D - \var[X] \left(1 + \tfrac{D}{\var[X]} + g(P)\right)^2} \right) \quad \text{if } (D,P) \in \mathcal{S}
        \end{cases}
    \end{split}\label{eq:RDP_GaussianKL}
\end{align}
{where 
\begin{align}
g(P) &= \frac{1}{\lamb[-1]\left( -e^{-(2P+1)} \right)}\label{RDP_DKL:g}\\
\mathcal{S} &= \left\{ (D,P): \frac{|\var[X] - D|}{\var[X]} \le -g(P) \right\}.\label{RDP_DKL:CaseIIIRegion}
\end{align}
Moreover, the solution in \eqref{eq:RDP_GaussianKL} is achieved by a linear realization $\hat{X} = aX + W$  such that the design variables $(a,\var[W])$ are obtained in closed-form as follows:} 
\begin{align}
    {a}&=
    \begin{cases}
        \max \left\{1 - \frac{D}{\var[X]}, 0 \right\} \quad \text{if } (D,P) {\in} \mathcal{S}^c\\
        \tfrac{1}{2}\left( 1 - \frac{D}{\var[X]} - g(P) \right) \quad \text{if } (D,P) \in \mathcal{S}
    \end{cases}\label{design_variables_KL_1}\\
    \var[W] &= D - \left( 1 - {a}\right)^2\var[X] \label{design_variables_KL_2}.
\end{align}
\end{theorem}

\begin{IEEEproof}
    See Appendix \ref{proof:uRDP:KL}.
\end{IEEEproof}

{Next, we use Theorem \ref{th:uRDP:KL} to obtain a similar result when the perception constraint in \eqref{opt: uRDPGaussian} is the reverse KL divergence. Our derivation extends \cite{qian:2023} providing, on top of a closed-form solution to the problem, the design of the optimal (forward) realization that achieves the optimal solution. This is reported in the following corollary.}

\begin{corollary}[{RDPF under reverse $D_{KL}(\cdot||\cdot)$}] \label{th:RDP:KLreverse}
{Consider the characterization in Problem \ref{problem:1} with $d(\cdot,\cdot)\equiv{D}_{KL}(\pdf{\hX}||\pdf{X})$. Then, the solution of \eqref{opt: uRDPGaussian} corresponds to $R(D,P)=\eqref{eq:RDP_GaussianKL}$, where in place of \eqref{RDP_DKL:CaseIIIRegion} we now have $g(P) = \lamb[0]\left( -e^{-(2P+1)} \right)$. Moreover, the optimal realization $\hat{X} = aX + W$ that achieves $R(D,P)$ corresponds to the choice of the design variables $(a,\var[W])$ given by \eqref{design_variables_KL_1} and \eqref{design_variables_KL_2}, with the updated form of $g(P)$ given above.} 
\end{corollary}
{\begin{IEEEproof}
The proof follows almost verbatim to the proof of Theorem \ref{th:uRDP:KL} hence we omit it.
\end{IEEEproof}
}

{In the next lemma, we provide an alternative derivation to a closed-form result first obtained by Zhang {\it et al.} in \cite[Theorem 1]{zhang:2021}. This result corresponds to the computation of \eqref{opt: uRDPGaussian} when $d(\cdot,\cdot)\equiv\WTS(\pdf{X}, \pdf{\hX})$. As in Corollary \ref{th:RDP:KLreverse}, we note that our methodology reveals on top of the closed-form solution to this problem, the design of the optimal (forward) realization that achieves the optimal solution.}
{These results are reported in the following lemma.
\begin{lemma} \label{th:uRDP:W2}
Consider the characterization in Problem \ref{problem:1} with $d(\cdot,\cdot)\equiv{W}_2^2(\pdf{X}||\cdot)$. Then the solution of \eqref{opt: uRDPGaussian} is obtained analytically as follows:
\begin{align}
    \begin{split}
        &R(D,P) = \\
        &\begin{cases}
                \max \left\{ \tfrac{1}{2} \log \left( \frac{\var[X]}{D} \right), 0 \right\}~~\text{if } (D,P) \in \mathcal{S}^c\\
                \tfrac{1}{2} \log \left( 1 + \tfrac{\var[X] + \left[(\stv[X] - \sqrt{P})^2 - D\right]^2}{(D - P) \left[(2\stv[X] - \sqrt{P})^2 - D\right]} \right)~~\text{if } (D,P) \in \mathcal{S}\\
        \end{cases} \label{eq:RDP_W2}
    \end{split}
\end{align}
where  $\mathcal{S} = \left\{ (D,P): \sqrt{P} \le \stv[X] - \sqrt{|\var[X] - D|} \right\}$. Moreover, the solution in \eqref{eq:RDP_GaussianKL} is achieved by a linear realization $\hat{X} = aX + W$  such that the design variables $(a,\var[W])$ are obtained in closed-form as follows:
\begin{align}
    a= 
    \begin{cases}
        \max \left\{1 - \frac{D}{\var[X]}, 0 \right\} \quad \text{if } (D,P) \in \mathcal{S}^c\\
        \tfrac{1}{2} \frac{\var[X] + \left(\stv[X] - \sqrt{P} \right)^2 - D}{\var[X]} \quad \text{if } (D,P) \in \mathcal{S}\\
    \end{cases},~~\var[W] =\eqref{design_variables_KL_2}.\label{design_variables_wesserstein}
\end{align}
\end{lemma}
\begin{IEEEproof}
    See Appendix \ref{proof:uRDP:W2}.
\end{IEEEproof}
}

{In the next theorem, we derive the closed-form solution of \eqref{opt: uRDPGaussian} when the perception constraint is the $\GJS$ divergence.}
\begin{theorem}[{RDPF under $\GJS(\cdot||\cdot)$}] \label{th:uRDP:GJS}
{Consider the characterization in Problem \ref{problem:1} with $d(\cdot,\cdot)\equiv{D}_{GJS}(\pdf{X}||\pdf{\hX})$. Then, the solution of \eqref{opt: uRDPGaussian} corresponds to the following analytical expression:
\begin{align} 
\begin{split}
R(D,P) = \begin{cases}
        \max \left\{ \tfrac{1}{2} \log \left( \frac{\var[X]}{D} \right), 0 \right\}~ ~\text{if } (D,P) \in \mathcal{S}^c\\
        \tfrac{1}{2} \log \left(1 + \frac{\var[X]\upsilon^2}{D - \var[X] \left(1 +\upsilon\right)^2} \right) ~~\text{if } (D,P) \in \mathcal{S}\\
\end{cases}
\end{split}\label{eq:RDP_GaussianGJS}
\end{align}
where $\upsilon=\tfrac{D}{2\var[X]} - \tfrac{g(P)}{4} + \tfrac{\sqrt{g(P)(g(P)-4)}}{4}$ with
\begin{align*}
g(P) &= -2 \lamb[-1]\left( - 2 e^{-(4P + 2)} \right)\\
    \mathcal{S}^c &= \left\{(D,P): f_L(P) \le |\var[X] - D| \le f_U(P) \right\}\\
    f_L(P)& = \tfrac{1}{2} \var[X] \left( -\sqrt{g(P)(g(P)-4)} - (2 - g(P))\right)\\ 
    f_U(P) &= \tfrac{1}{2} \var[X] \left( \sqrt{g(P)(g(P)-4)} - (2 - g(P))\right). 
\end{align*} 
Moreover, the solution in \eqref{eq:RDP_GaussianGJS} is achieved by a linear realization $\hat{X} = aX + W$ such that the design variables $(a,\var[W])$ are obtained in closed-form as follows:
\begin{align}
a&=\begin{cases}
        \max \left\{1 - \frac{D}{\var[X]}, 0 \right\} \quad \text{if } (D,P) \in \mathcal{S}^c\\
        \tfrac{1}{2}\left( 1 - \frac{D - f_L(P)}{\var[X]} \right) \quad \text{if } (D,P) \in \mathcal{S}\\
    \end{cases},~~\var[W] =\eqref{design_variables_KL_2}.\label{design_variables_GJS}
\end{align}}
\end{theorem}
\begin{IEEEproof}
See Appendix \ref{proof:uRDP:GJS}.
\end{IEEEproof}

{Our last result in this section provides the closed-form solution of \eqref{opt: uRDPGaussian} when the perception constraint is the $\HS$ distance.}
\begin{theorem}[{RDPF under $\HS(\cdot,\cdot)$}] \label{th:uRDP:HS}
{Consider the characterization in Problem \ref{problem:1} with $d(\cdot,\cdot)\equiv \HS(\pdf{X},\pdf{\hX})$. Then, the solution of \eqref{opt: uRDPGaussian} corresponds to the following analytical expression:
\begin{align} 
\begin{split}
R(D,P) = \begin{cases}
        \max \left\{ \tfrac{1}{2} \log \left( \frac{\var[X]}{D} \right), 0 \right\}~ ~\text{if } (D,P) \in \mathcal{S}^c\\
        \tfrac{1}{2} \log \left(1 + \frac{ \var[X]\left(g(P) - \frac{D}{2 \var[X]} \right)^2}{D - \var[X] \left(1 - g(P) + \frac{D}{2 \var[X]} \right)^2 } \right) ~~\text{if } (D,P) \in \mathcal{S}\\
\end{cases}
\end{split}\label{eq:RDP_GaussianHS}
\end{align}
where 
\begin{align*}
    g(P) &= \frac{1 - \sqrt{1 - (1-\frac{P}{2})^4}}{(1-\frac{P}{2})^4}\\
    \mathcal{S} &= \left\{ (D,P): 1-g(P) \le \frac{D}{2\var[X]} \le g(P) \right\}. 
\end{align*} 
Moreover, the solution in \eqref{eq:RDP_GaussianHS} is achieved by a linear realization $\hat{X} = aX + W$ such that the design variables $(a,\var[W])$ are obtained in closed-form as follows:
\begin{align}
a&=\begin{cases}
        \max \left\{1 - \frac{D}{\var[X]}, 0 \right\} \quad &\text{if } (D,P) \in \mathcal{S}^c\\
        g(P) - \frac{D}{2\var[X]} \quad  &\text{if } (D,P) \in \mathcal{S}\\
    \end{cases},~~\var[W] =\eqref{design_variables_KL_2}.\label{design_variables_HS}
\end{align}}
\end{theorem}
\begin{IEEEproof}
See Appendix \ref{proof:uRDP:HS}.
\end{IEEEproof}

\section{RDPF for Vector-Valued Gaussian sources} \label{sec: MultivariateGaussianSourcesRDPF}
{In this section, our goal is twofold. First, we derive (under certain conditions) a generic alternating minimization approach and its algorithmic embodiment, which allows the computation of any RDPF for multidimensional Gaussian sources. Second, we apply this algorithm when having an MSE distortion constraint and the class of perception constraints studied in Section \ref{sec: unvariate closed forms}, namely $\KL,~\GJS,~\HS,~\WTS$.}

{\subsection{A Generic Alternating Minimization Approach}\label{subsec:generic_algo}}

{We start with the generalization of Problem \ref{problem:1} to vector-valued Gaussian sources. In contrast to Problem \ref{problem:1} we do not specify the type of fidelity constraints that will be utilized. This is because our aim is to first provide a general approach along with its algorithmic embodiment, able to tackle \emph{any} vector-valued jointly Gaussian problems for RDPF.
\begin{problem}\label{problem:2}
Given a Gaussian source $X \sim \ND(\mean[X], \Cov[X])$, $\Cov[X] \succ 0$, assume that the reconstructed random vector $\hat{X}\in\mathbb{R}^N$ is chosen such that the joint tuple $(X,\hat{X})$ is jointly Gaussian. Then, the reconstructed message admits a linear (forward) realization of the form $\hat{X} = AX + W$, where $A\in\mathbb{R}^{N\times{N}}$, $W \sim \ND(\mean[W], \Cov[W])$, $\Cov[W] \succeq 0$, such that $\mean[\hX] = A\mean[X] + \mean[W]$ and $\Cov[\hX] = A\Cov[X]A^T + \Cov[W]$. By considering the case where $\mean[X]=\mean[\hX]$ (by setting $\mean[W] = (I - A)\mean[X]$), we can cast \eqref{opt:RDPF}-\eqref{opt: perception_constraint}  as follows:
\begin{align}
    \begin{split}
        R(D,P) &=  \min_{A\in\mathbb{R}^{N\times{N}},\Cov[W]\succeq{0}} \tfrac{1}{2} \log \left(\tfrac{|A\Cov[X]{A}^T + \Cov[W]|}{|\Cov[W]|} \right)\\
        \textrm{s.t.}   & \quad \E{\Delta(X,\hX)} \le D \\
                    & \quad d(\pdf{X}||\pdf{\hX}) \le P
    \end{split} \label{opt: generic_gaussian_vector}
\end{align}
where the specific form of the fidelity constraints $\E{\Delta(\cdot,\cdot)}$ and $d(\cdot,\cdot)$ are not yet specified.
\end{problem}
}

Assuming that also the fidelity constraints $\E{\Delta(\cdot,\cdot)}$ and $d(\cdot,\cdot)$ are tensorizable, i.e., 
\begin{align*}
    \E{\Delta(X,\hX)} &\ge \sum_{i = 1}^N g\left(\E{\Delta(X_i,\hX_i)}\right)\\
    d(\pdf{X},\pdf{\hX}) &\ge \sum_{i = 1}^N h\left(d(\pdf{X_i},\pdf{\hX_i})\right)
\end{align*}
with $g(\cdot)$ and $h(\cdot)$ convex functions, then {applying [Appendix \ref{sec:tensorization}, Lemma \ref{th:gmrdpf:MI_Commutation}] in \eqref{opt: generic_gaussian_vector} leads to the following lower bound:
\begin{align}
    \begin{split}
        R(D,P) \stackrel{(\star)}\ge & \min_{\eigvi{A}, \eigvi{\Cov[W]}} \sum_{i=1}^N \tfrac{1}{2}\log\left(1+\tfrac{\eigvi{A}^2\eigvi{X}}{\eigvi{W}}\right)\\
        \textrm{s.t.}   & \quad \sum_{i=1}^N g\left(\E{\Delta(X_i,\hat{X}_i)}\right) \le D \\
                        & \quad \sum_{i=1}^N h\left(d(\pdf{X_i}||\pdf{\hat{X}_i})\right) \le P 
   \end{split} \label{opt: RDPFonTheEigenvectors}
\end{align}
where $(\star)$ holds with equality if the triplet $(A,\Cov[W], \Cov[X])$ commute by pairs \cite[Section 0.7.7]{horn:2012:matrix}. In fact, for jointly Gaussian random vectors $(X,\hX)$, the sufficient condition that achieves the lower bound in \eqref{opt: RDPFonTheEigenvectors} is always true\footnote{This is because the matrices $(A, \Cov[W])$ are design variables and can be chosen such that they have the \textit{same eigenvectors} as $\Cov[X]$, for details see, e.g., \cite[Proposition 1]{stavrou:2022}.}, hence one can replace inequality with equality  without loss of generality.} 
{ \paragraph*{Proposed alternating minimization method} To solve \eqref{opt: RDPFonTheEigenvectors}, we first introduce the (vector) optimization variables $\vecD = [D_i]_{i \in 1:N}$ and $\vecP = [P_i]_{i \in 1:N}$ such that
\begin{align*}
D_i = \E{\Delta(X,\hat{X}_i)},~~P_i = d(\pdf{X_i}||\pdf{\hat{X}_i}) \quad\forall i \in 1:N.
\end{align*}
Once the slack variables above are substituted in \eqref{opt: RDPFonTheEigenvectors}, we yield:
\begin{align}
    \begin{split}
        R(D,P) = & \min_{\vecD, \vecP} \sum_{i=1}^N R_i(D_i,P_i) \\
        \textrm{s.t.}   & \quad \sum_{i=1}^N g(D_i) \le D \qquad \sum_{i=1}^N h(P_i) \le P 
   \end{split} \label{opt:gmrdpf:full}
\end{align}
where $R_i(D_i,P_i)$ corresponds to the stagewise RDPF given by
\begin{align}
   \begin{split}
       R_i(D_i,P_i) = & \min_{\eigvi{A}, \eigvi{\Cov[W]}} \tfrac{1}{2}\log\left(1+\tfrac{\eigvi{A}^2\eigvi{X}}{\eigvi{W}}\right).  \\
        \textrm{s.t.}   
        & \quad D_i=\E{\Delta(X_i,\hat{X}_i)}  \\
        & \quad P_i=d(\pdf{X_i}||\pdf{\hat{X}_i}). 
   \end{split}
\end{align}
Note that in \eqref{opt:gmrdpf:full} we have three distinct ``rate region'' cases, which combined, result into the complete RDPF. Particularly, we may have only the distortion constraint to be active (hereinafter referred to as \textbf{Case I}), only the perception constraint to be active (hereinafter referred to as \textbf{Case II}), or both constraints to be active (hereinafter referred to as \textbf{Case III}). We note that only \textbf{Case III} is interesting as the computation of the other two cases follows from the computation of that case.}
\par {To find the optimal pair $(\vecD^{*},\vecP^{*})$ in \eqref{opt:gmrdpf:full} we resort to an application of an alternating minimization technique. Specifically, we define the following two subproblems of \eqref{opt:gmrdpf:full}:
\begin{itemize}
    \item For fixed $\vecP$, \eqref{opt:gmrdpf:full} simplifies to
        \begin{align}
            \begin{split}
               \min_{\vecD} \sum_{i=1}^N R_i(D_i,P_i) \quad \textrm{s.t.} \quad \sum_{i=1}^N g(D_i) \le D. \\
           \end{split} \label{opt:gmrdpf:partial_optD}
        \end{align} 
    \item For fixed $\vecD$, \eqref{opt:gmrdpf:full} simplifies to
    \begin{align}
        \begin{split}
           \min_{\vecP} \sum_{i=1}^N R_i(D_i,P_i)  \quad \textrm{s.t.}  \quad \sum_{i=1}^N h(P_i) \le P. \\
       \end{split} \label{opt:gmrdpf:partial_optP}
    \end{align}
\end{itemize}
It should be noted that solving the optimization problems of \eqref{opt:gmrdpf:partial_optD} and \eqref{opt:gmrdpf:partial_optP} is of primary interest because the solution of these two problems forms the basis of an alternating minimization scheme that can optimally solve \eqref{opt:gmrdpf:full}. To make this point clear, next, we state and prove the convergence to an optimal point of an alternating minimization approach that relies on the solution of \eqref{opt:gmrdpf:partial_optD} and \eqref{opt:gmrdpf:partial_optP}. This is often encountered in the literature as {\it block nonlinear Gauss–Seidel method} \cite{grippo:2000}.
\begin{theorem}(Convergence) \label{th:gmrdpf:RDPF_AM}
    Let the optimization problem \eqref{opt:gmrdpf:full} be defined for finite distortion and perception levels $(D,P)$. Let  $(\vecD^{(0)},\vecP^{(0)})$ be an initial point and let the sequence $\{(\vecD^{(n)},\vecP^{(n)}):~n=1,2,\ldots\}$ be the sequence obtained by the alternating optimization of problems \eqref{opt:gmrdpf:partial_optD} and \eqref{opt:gmrdpf:partial_optP}. Then the sequence has a limit $\lim_{n \to \infty}(\vecD^{(n)},\vecP^{(n)})  = (\vecD^{*},\vecP^{*})$ and the limit is an optimal solution of \eqref{opt:gmrdpf:full}. 
\end{theorem}
\begin{IEEEproof}
    See Appendix \ref{proof:gmrdpf:RDPF_AM}.
\end{IEEEproof}
}
\begin{remark}
The assumption of finite $(D,P)$ levels in Theorem \ref{th:gmrdpf:RDPF_AM} implies that the Lagrangian multipliers $(s_1, s_2)$ associated with the distortion and perception constraints are strictly positive. However, under the assumption of bounded perception metric $d(\cdot,\cdot)$, the case $s_2 = 0$ does not violate the assumptions of Theorem \ref{th:gmrdpf:RDPF_AM}. In this regime, the perception constraint of Problem \ref{problem:2} is inactive, making the problem equivalent to the classical RDF problem.
\end{remark}

\begin{algorithm}
    \caption{Algorithm of Theorem \ref{th:gmrdpf:RDPF_AM}} \label{alg: AM}
    \begin{algorithmic}[1]
        
        \Require source distribution $\pdf{X} = \ND(\mu_X, \Cov[X])$ with $\Cov[X] \succ 0$; Lagrangian parameters $s = (s_1, s_2)$ with $s_1 > 0$ and $s_2 > 0$; error tolerances $\epsilon$; initial point $(\vecD^{(0)},\vecP^{(0)})$.
        \item[]
        \State  $\omega \gets +\infty$; $n \gets 1$;
        
        \While{$\omega > \epsilon$}
            \State $\vecD^{(n)} \gets \text{Solution Problem } \eqref{opt:gmrdpf:partial_optD} \text{ for } (\vecP^{(n-1)}, s_1)$

            \State $\vecP^{(n)} \gets \text{Solution Problem } \eqref{opt:gmrdpf:partial_optP} \text{ for } (\vecD^{(n)}, s_2)$
            
            \State $\omega \gets || (\vecD^{(n)},\vecP^{(n)}) - (\vecD^{(n-1)},\vecP^{(n-1)})||_2$ 
        
        \State $n \gets n + 1$

        \EndWhile
        \item[]
    \Ensure { $D = \sum_{i=1}^N g(D_i^{(n)}), ~ P = \sum_{i=1}^N h(P_i^{(n)})$, ~$R(D,P) = \sum_{i=1}^N R_i(D_i^{(n)}, P_i^{(n)})$.}

    \end{algorithmic}
\end{algorithm}
{In Algorithm \ref{alg: AM} we implement the alternating minimization scheme of Theorem \ref{th:gmrdpf:RDPF_AM}, which allows for the computation of any multivariate Gaussian RDPF of the form characterized in Problem \ref{problem:2} as long as we can have a characterization of the problem for the univariate case. In other words, we can always cast the general Gaussian problem into the optimization problem in \eqref{opt:gmrdpf:full}, which further means that the proposed alternating minimization approach can be applied whenever the  RDPF can be obtained in closed-form for scalar-valued Gaussian sources.} 
\par {The computation of the optimal pair $(\vecD^*,\vecP^*)$ via Algorithm \ref{alg: AM}, can be used to obtain numerically the value of matrices $(A,\Cov[W])$. This means that we can numerically compute the linear realization $\hat{X} = AX + W$ that achieves Problem \ref{problem:2}. On top of that, by leveraging the fact that $(A,\Cov[W],\Cov[X])$, commute by pairs, we can expect that the aforementioned design variables will be of the form:
\begin{align}
    A&=V \cdot \diag\left( \left[\eigvi{A}\right]_{i=1:N}\right) \cdot V^T,\nonumber\\
    \Cov[W]&= V \cdot \diag\left( \left[\eigvi{\Cov[W]}\right]_{i=1:N} \right) \cdot V^T\nonumber
\end{align}
where $V\in\mathbb{R}^{N\times{N}}$ is a non-singular matrix, and $\{(\eigvi{A}, \eigvi{\Cov[W]}):~i=1,2,\ldots,N\}$ are both functions of $\{(\eigvi{X}, D_i, P_i):~i=1,2,\ldots,N\}$ which are associated with the parameters for the associated stagewise solution of RDPF.}

\paragraph*{Analysis of Algorithm \ref{alg: AM}} To characterize the worst-case performance of Alg. \ref{alg: AM}, we provide an upper bound on its convergence rate.

\begin{theorem}(Upper bound on the Convergence Rate) \label{th:conv:convrate}
Let $R(\cdot,\cdot)$ be the function defined in \eqref{opt:gmrdpf:full} and let $s_1 >0$, $s_2>0$ be the Lagrangian multipliers associated with the target distortion level $D$ and perception level $P$. Let $(\vecD^{(n)},\vecP^{(n)})$ for $n = 0,\dots,T$ be the sequence of iterations generated by Alg. \ref{alg: AM}. Then, the rate of convergence to the stationary point is upper bounded by $\mathcal{O}\left(\frac{1}{\sqrt{T}}\right)$ (sublinear convergence rate).   
\end{theorem}
\begin{IEEEproof}
    See Appendix \ref{proof:conv:convrate}.
\end{IEEEproof}

In what follows, we demonstrate experimentally that indeed at worse Alg. \ref{alg: AM} can achieve sublinear convergence, but depending on the value of the Lagrangian multipliers the performance can significantly improve, achieving linear convergence rate. 

\begin{example} \label{example:experimantal_convergence}
    Let $X \sim \ND(0,\Cov[X])$ with $\Cov[X] = \diag([1,3,5,7,10])$ and let Problem \ref{problem:2} be defined for MSE distortion metric and $\WTS$ perception metric. Then, Figure \ref{fig: experimental_convergence} shows the convergence rate of Alg. \ref{alg: AM} for different values of the Lagrangian multipliers $s_1 \in \{ 10^{-1},10^{-2},10^{-3},10^{-4}\}$ and $s_2 \in \{ 1,10^{-1},10^{-2},10^{-3},10^{-4}\}$. Clearly, Fig. \ref{fig: experimental_convergence} illustrates that depending on the values of the Lagrangian multipliers, the convergence rate of the algorithm can be improved, achieving even linear convergence rates.

    \begin{figure}[htbp]
        \centering
        \includegraphics[width = \linewidth]{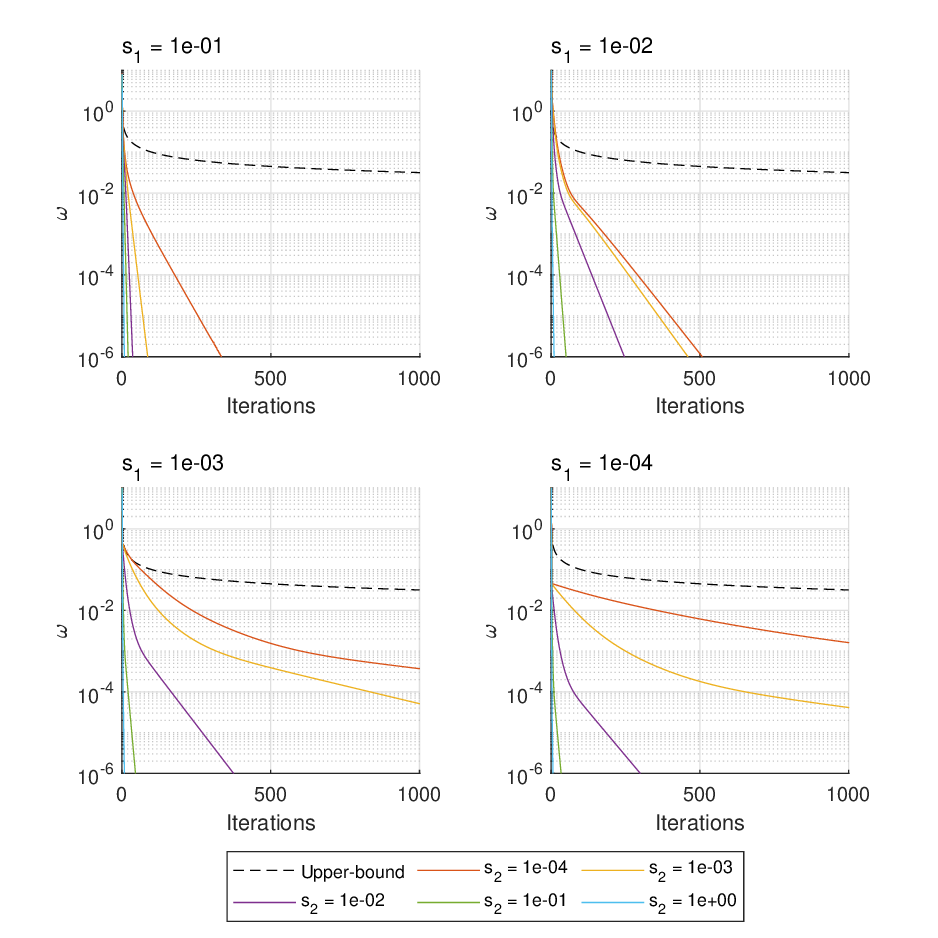}
        \caption{Experimental convergence rate of Alg. \ref{alg: AM} for a 5-dimensional Gaussian source.}
        \label{fig: experimental_convergence}
    \end{figure}
\end{example}

\subsection{Application of the Alternating Minimization Approach} \label{subsec:applicationAlg}

\begin{table*}[htp]
    \caption{Closed-Form Solutions of Subproblem \eqref{opt:gmrdpf:partial_optD}}
    \label{table:subDClosedForm}
    \begin{tabular}{ l c }
        \toprule
        $d(\pdf{X},\pdf{\hat{X}})$  & $\vecD^*$ \\ 
        \midrule
        $\KL(\pdf{X}||\pdf{\hat{X}})$   & $D^*_i = \frac{1}{2s_1} + \eigvi{\Cov[X]}\left(1 - \frac{1}{\lamb[-1]\left( - e^{-(2P_i +1) }\right)} \right) +  \frac{1}{\lamb[-1]\left( - e^{-(2P_i +1) }\right)} \sqrt{\lamb[-1]\left( - e^{-(2P_i +1) }\right)\left( \frac{\lamb[-1]\left( - e^{-(2P_i +1) }\right)}{4s_1^2} -4\eigviSq{\Cov[X]}\right)}$   \\
        \\
        $\KL(\pdf{\hat{X}}||\pdf{X})$   & $D^*_i = \frac{1}{2s_1} + \eigvi{\Cov[X]}\left(1 - \lamb[0]\left( - e^{-(2P_i +1) }\right)\right) - \sqrt{4\eigviSq{\Cov[X]}(-\lamb[0]\left( - e^{-(2P_i +1) }\right)) + \frac{1}{4s_1^2}}$   \\
        \\
        \multirow{1}{*}{$\HS(\pdf{X}||\pdf{\hat{X}})$}  & $D^*_i = \frac{1}{2s_1} + 2Z(P_i)\eigvi{\Cov[X]}  - \sqrt{\frac{1}{4s_1^2} + 4\eigviSq{\Cov[X]}\left(2Z(P_i) - 1\right)},~\quad Z(P_i) = \frac{1 - \sqrt{1 - (1-\frac{P_i}{2})^4}}{(1-\frac{P_i}{2})^4}$\\
        \\
        \multirow{2}{*}{$\GJS(\pdf{X}||\pdf{\hat{X}})$}  & $D^*_i = \frac{1}{2s_1} -\eigvi{\Cov[X]}\left(Z(P_i) + \sqrt{Z(P_i)(Z(P_i) + 2)}\right)  - \sqrt{\frac{1}{4s_1^2} - 4\eigviSq{\Cov[X]}\left(Z(P_i) + \sqrt{Z(P_i)(Z(P_i) + 2)} + 1\right)}$, \\
            & $Z(P_i) = \lamb[-1]\left( - e^{-(4P_i + 2) }\right)$\\
        \bottomrule
    \end{tabular}
\end{table*}

\par {In this subsection, we specialize the analysis of Subsection \ref{subsec:generic_algo} to the MSE distortion constraint whereas as perception constraint we consider the squared Wasserstein-2 distance. We remark that similar specializations can be developed for other divergence measures, under the assumption that a closed-form solution for scalar-valued Gaussian source RDPF is available (see e.g., Section \ref{sec: unvariate closed forms}).}
We start by solving the subproblems \eqref{opt:gmrdpf:partial_optD} and \eqref{opt:gmrdpf:partial_optP}. To this end, we leverage Lemma \ref{th:uRDP:W2} to characterize the function $R_i(D_i,P_i)$, which is the stagewise RDPF for the $i^{th}$ dimension, under $\MSE$ distortion metric and $\WTS$ perception metric, for a Gaussian source $X_i \sim \ND(0,\eigvi{\Cov[X]})$. Furthermore, Lemma \ref{th:uRDP:W2} and Proposition \ref{th:gmrdpf:W2_Commutation} defines the functional form of as the auxiliary optimization variables $\vecD = [D_i]_{i \in 1:N}$ and $\vecP = [P_i]_{i \in 1:N}$ and of the tensorization functions $g(\cdot)$ and $h(\cdot)$, introduced in \eqref{opt:gmrdpf:full}, as follows:
\begin{align*}
    D_i &= \E{||X_i - \hX_i||^2} = (1 - \eigvi{A})^2\eigvi{\Cov[X]} + \eigvi{\Cov[W]}\\
    P_i &= \WTS(\pdf{X_i},\pdf{\hX_i}) =\eigvi{\Cov[X]} - \sqrt{\eigvi{A}^2 \eigvi{\Cov[X]} + \eigvi{\Cov[W]}}\\
    g(\cdot) &= h(\cdot) =  \id(\cdot)
\end{align*}
where $\hX_i \sim \ND(0, \eigvi{\Cov[X_i]})$ is the stagewise linear realization of the form $\hX_i = \eigvi{A} X_i + W_i$ with $W_i \sim \ND(0, \eigvi{\Cov[W]})$.
\par In the following theorem we derive the characterization of the optimal solution of the subproblem \eqref{opt:gmrdpf:partial_optD}.

\begin{theorem} \label{th:gmrdpf:AM_subD}
    Let the Lagrangian multiplier $s_1 > 0$ be given. Then \eqref{opt:gmrdpf:partial_optD} can be expressed as follows:
    \begin{align*}
        \tilde{R}(\vecP) = -s_1D + \min_{\vecD} \sum_{i = 1}^N R_i(D_i,P_i) + s_1 \sum_{i = 1}^N D_i
    \end{align*}
    where $D = \sum_{i = 1}^N D_i^*$ with $\vecD^* = [D_i^*]_{i \in 1:N}$ being the optimal stagewise distortions levels achieving the minimum. Furthermore, $D^*_i(\vecP) \in \mathcal{S}$ is characterized as:
    \begin{align}
        \begin{split}
            D^*_i &=  P_i + 2 \stveigvi[X] \left(\stveigvi[X] - \sqrt{P_i} \right) \\
            &\quad + \left(\frac{1}{2s_1} - \sqrt{4\eigvi{\Cov[X]}(\stveigvi[X] - \sqrt{P_i})^2 + \frac{1}{4s_1^2}} \right). 
        \end{split} \label{eq:gmrdpf:AM_subD:optD}
    \end{align}
\end{theorem}
\begin{IEEEproof}
    See Appendix \ref{proof:gmrdpf:AM_subD}.
\end{IEEEproof}

We remark that similar expressions to \eqref{eq:gmrdpf:AM_subD:optD} in Theorem \ref{th:gmrdpf:AM_subD} can be derived for other perception constraints. Such cases are reported in Table \ref{table:subDClosedForm}.
We now move to the characterization of the optimal solution of the subproblem \eqref{opt:gmrdpf:partial_optP}.

\begin{theorem} \label{th:gmrdpf:AM_subP}
    Let the Lagrangian multiplier $s_2 > 0$ be given. Then \eqref{opt:gmrdpf:partial_optP} can be expressed as follows:
    \begin{align*}
        \hat{R}(\vecD) = -s_2P + \min_{\vecP} \sum_{i=1}^N R_i(D_i,P_i) + s_2\sum_{i=1}^N P_i
    \end{align*}
    where $P = \sum_{i=1}^N P_i^*$ with $\vecP^* = [P_i^*]_{i \in 1:N}$ being the optimal stagewise perception levels achieving the minimum. Furthermore, the components of the optimal solution $\vecP^*$ can be characterized as the zeros of the vector function $T(\cdot):\mathbb{R}^N \to \mathbb{R}^N$ where each component is defined as:
    \begin{align}
        T_i(x) = \frac{\partial R_i(D_i,P_i)}{\partial P_i}\Bigg|_{x_i} + s_2 \label{eq:solsubP:T}
    \end{align}
    
    \begin{align}
        \begin{split}
            &\frac{\partial R_i(D_i,P_i)}{\partial P_i} =\\ 
            &\begin{cases}
                \tfrac{1}{2} \frac{(\stveigvi[X] - \sqrt{P_i})^4 - ( \eigvi{\Cov[X]} - D_i)^2}{\sqrt{P_i}(D_i - P_i)(\sqrt{P_i} - \stveigvi[X])(D_i - (2\stveigvi[X] - \sqrt{P_i})^2)} &\text{if } P_i \in \mathcal{S}\\\\
                0 & \text{if } P_i \in \mathcal{S}^c.
            \end{cases}
        \end{split}
    \end{align}
\end{theorem}
\begin{IEEEproof}
    See Appendix \ref{proof:gmrdpf:AM_subP}.
\end{IEEEproof}

\begin{remark}
    Similar expressions to \eqref{eq:solsubP:T} can be derived for the direct $\KL$, reverse $\KL$ and $\GJS$ divergences cases. On the contrary, the case of the $\HS$ distance requires particular care. Using Proposition \ref{th:gmrdpf:HS_Commutation}, we can express the $\HS$ tensorization inequality as
    \begin{align*}
        \HS(\pdf{X},\pdf{\hX}) \ge h^{-1}\left( \sum_{i=1}^N h \left(\HS(\pdf{X_i},\pdf{\hX_i})\right)\right)
    \end{align*}
    where $h:[0,2) \to \mathbb{R}^+$ is the convex strictly increasing bijection defined as $h(x) = -\log(1 - \frac{x}{2})$ and $h^{-1}$ is its inverse. Therefore, the perception constraint in \eqref{opt: RDPFonTheEigenvectors} can be expressed as follows:
    \begin{align*}
        h^{-1}\left( \sum_{i=1}^N h \left(\HS(\pdf{X_i},\pdf{\hX_i})\right)\right) \le P
    \end{align*}
    which, due to the monotonicity of $h$, is equivalent to
    \begin{align*}
        \sum_{i=1}^N h \left(\HS(\pdf{X_i},\pdf{\hX_i})\right) \le h(P).
    \end{align*}
    Using the modified perception level $P' = h(P)$, the new formulation respects the constraint format described in \eqref{opt:gmrdpf:full}.
\end{remark}

\begin{corollary}\label{lemma:PropertiesTi}
    Let $T_i:\mathbb{R}^N\to\mathbb{R}$ be the $i^{th}$ component of the vector function $T(\cdot)$ defined in Theorem \ref{th:gmrdpf:AM_subP}. Then, $T_i$ is a continuous and non-decreasing function on $\mathbb{R}$. Furthermore, $T_i$ has at least one root in $\mathcal{S}$.
\end{corollary}
\begin{IEEEproof}
    See Appendix \ref{proof:PropertiesTi}.
\end{IEEEproof}

Although we are not able to derive a closed-form solution for subproblem \eqref{opt:gmrdpf:partial_optP}, the optimal $\vecP^*$ can be found as zeros of the functions $\{T_i\}_{i\in1:N}$. Corollary \ref{lemma:PropertiesTi} guarantees that the functions $\{T_i\}_{i\in1:N}$ respect the assumptions required for the application of the bisection method \cite[Chapter 2.1]{burden:2015}. More refined root-finding methods, such as the Newton Method \cite[Chapter 2.3]{burden:2015}, are not applicable in this instance given the requirement on the differentiability of $T_i$, which cannot be guaranteed, in general.

\subsection{RDPF under "Perfect Realism" regime} \label{subsec: perfect_realism}

The results of Theorem \ref{th:gmrdpf:AM_subD} characterize the optimal distortion vector $\vecD$ for a given perception vector $\vecP$. As an additional result, Theorem \ref{th:gmrdpf:AM_subD} gives the closed form solution for the case where $\vecP = \mathbf{0}$, or equivalently $d(\pdf{X},\pdf{\hX}) = 0$, referred to as {\it perfect realism} \cite{chen:2022, wagner:2022:rate }.

\begin{corollary} \label{th: perfect_realism}
    Consider the optimization problem \eqref{opt:gmrdpf:full} for perception lever $P = 0$. Then, for a given Lagrangian multiplier $s_1 > 0$, the optimal solution $\vecD^* = [D_i^*]_{i \in 1:N}$ is given by
    \begin{align}
        D^*_i = 2 \eigvi{\Cov[X]} + \frac{1}{2s_1} - \sqrt{4\eigviSq{\Cov[X]} + \frac{1}{4s_1^2}} \label{eq:perfect_realism_solution}
    \end{align} 
   such that the distortion level $D = \sum_{i = 1}^N D^*_i$.
\end{corollary}
\begin{IEEEproof}
     \eqref{eq:perfect_realism_solution} is obtained from \eqref{eq:gmrdpf:AM_subD:optD} for $P_i = 0$. 
\end{IEEEproof}
We stress the following two technical remarks for Corollary \ref{th: perfect_realism}.
\begin{remark} \label{remark: perfect_realism_limits}
    The optimal solution $\vecD^*$ is well defined in the limit $s_1 \to 0$, since $\lim_{s_1 \to 0} D^*_i = 2 \eigvi{\Cov[X]}$.
\end{remark}

\begin{remark} \label{remark: comparison with WF}
In the water-filling solution of the classical multivariate Gaussian RD, the optimal solution $\vecD^*_{RD} = [D^*_{i,RD}]_{i \in 1:N}$ for a Lagrangian multiplier $s_1 > 0$  is given by
\begin{align*}
    D^*_{i,RD} = \min\left( w(s_1), \eigvi{\Cov[X]} \right) \qquad w(s_1) = \frac{1}{2s_1}
\end{align*}
where water-level $w(s_1)$ is dimension independent and the $\min(\cdot)$ operation is required to guarantee that $D^*_{i,RD}$ belongs to the constraint set.  Heuristically, one can imagine that the $i$-th source component is discarded in the reconstruction whenever $w(s_1) \ge \eigvi{\Cov[X]}$, upper bounding the maximum distortion observed in the $i$-th component by $\eigvi{\Cov[X]}$.
Conversely, the solution identified in \eqref{eq:perfect_realism_solution}, and in general the results of Theorem \ref{th:gmrdpf:AM_subD}, can be interpreted as an adaptive water-level. In fact, in \eqref{eq:perfect_realism_solution}, $w(s_1)$ gets adapted to each dimension, guaranteeing that all source components are present in the reconstructed signal. However, as already observed in \cite[Theorem 2]{blau:2019}, \eqref{eq:perfect_realism_solution} suggests that the maximum distortion on the $i$-th dimension is greater that $\eigvi{\Cov[X]}$ and is upper bounded by $2\eigvi{\Cov[X]}$.
\end{remark}

{\section{Numerical Simulations} \label{sec: numerical examples}}
{In this section, we provide numerical simulations based on the theoretical findings of Sections \ref{sec: unvariate closed forms}, \ref{sec: MultivariateGaussianSourcesRDPF}.}

\subsection{Scalar-valued Gaussian Sources}
\begin{figure}[t]
    \centering
    \begin{subfigure}{\linewidth}
		\centering \includegraphics[width=0.9\linewidth]{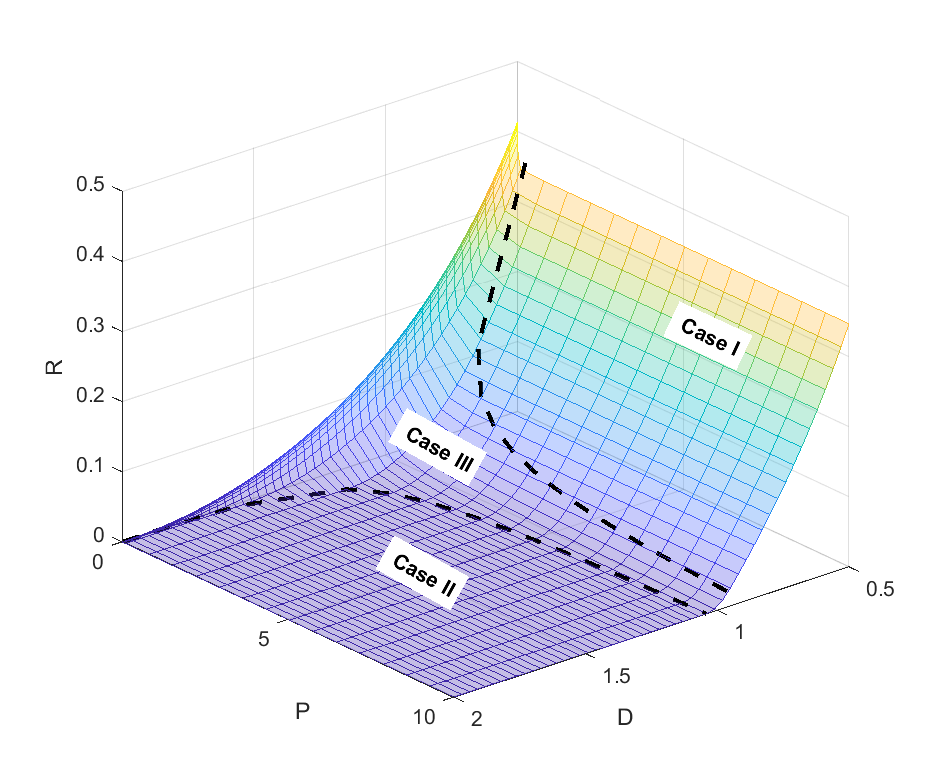}
		\caption{} \label{fig: uRDPKLDirect}
    \end{subfigure}\\
    \begin{subfigure}{0.49\linewidth}
        \centering \includegraphics[width=0.9\linewidth]{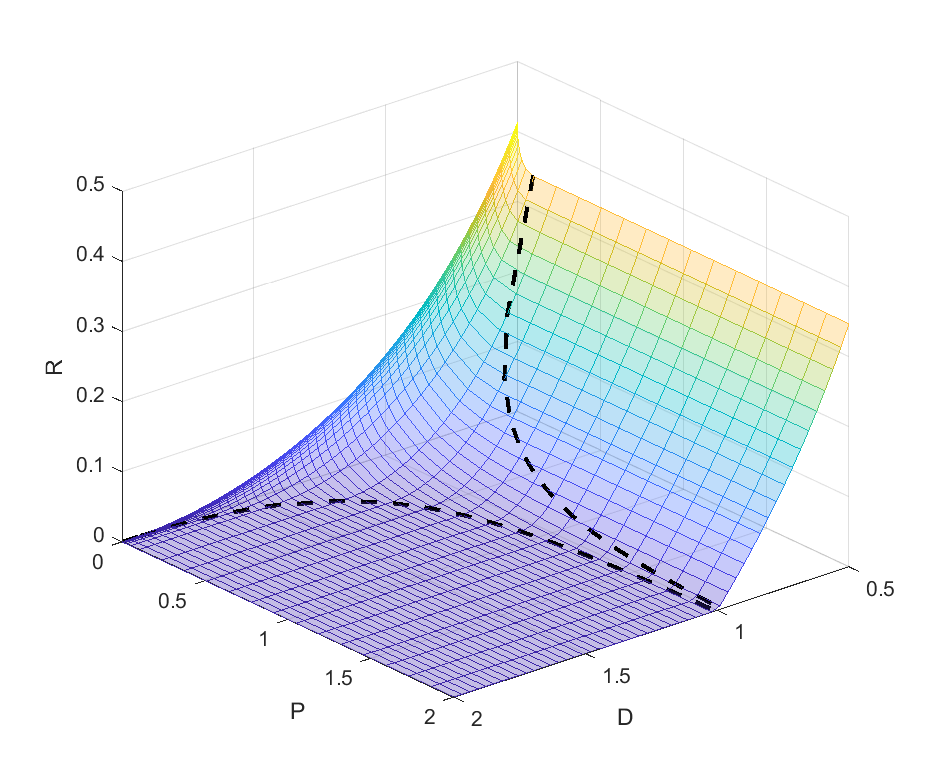}
        \caption{} \label{fig: uRDPKLReverse}
    \end{subfigure}
        \begin{subfigure}{0.49\linewidth}
        \centering \includegraphics[width=0.9\linewidth]{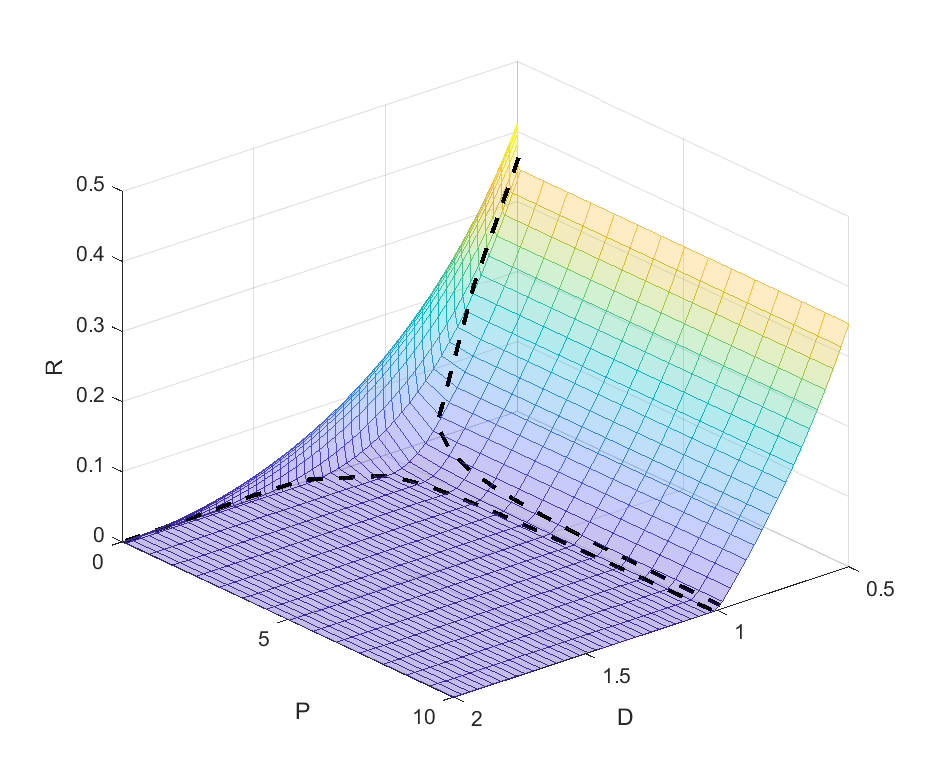}
        \caption{} \label{fig: uRDPGJS}
    \end{subfigure}\\
    \begin{subfigure}{0.49\linewidth}
        \centering \includegraphics[width=0.9\linewidth]{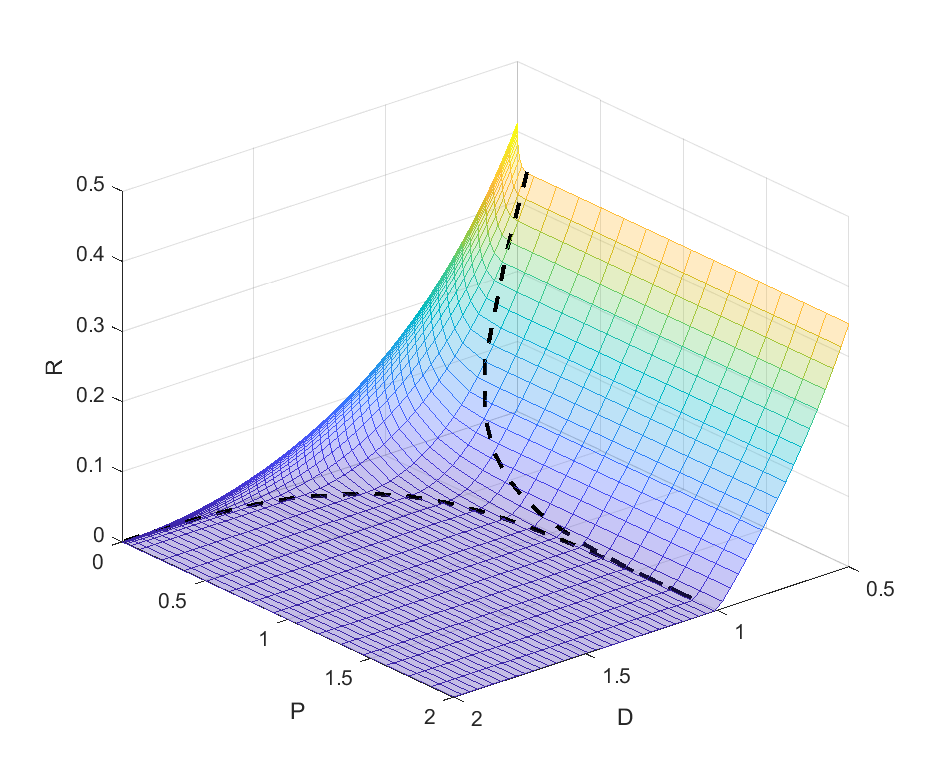}
        \caption{} \label{fig: uRDPHS}
    \end{subfigure}
    \begin{subfigure}{0.49\linewidth}
        \centering \includegraphics[width=0.9\linewidth]{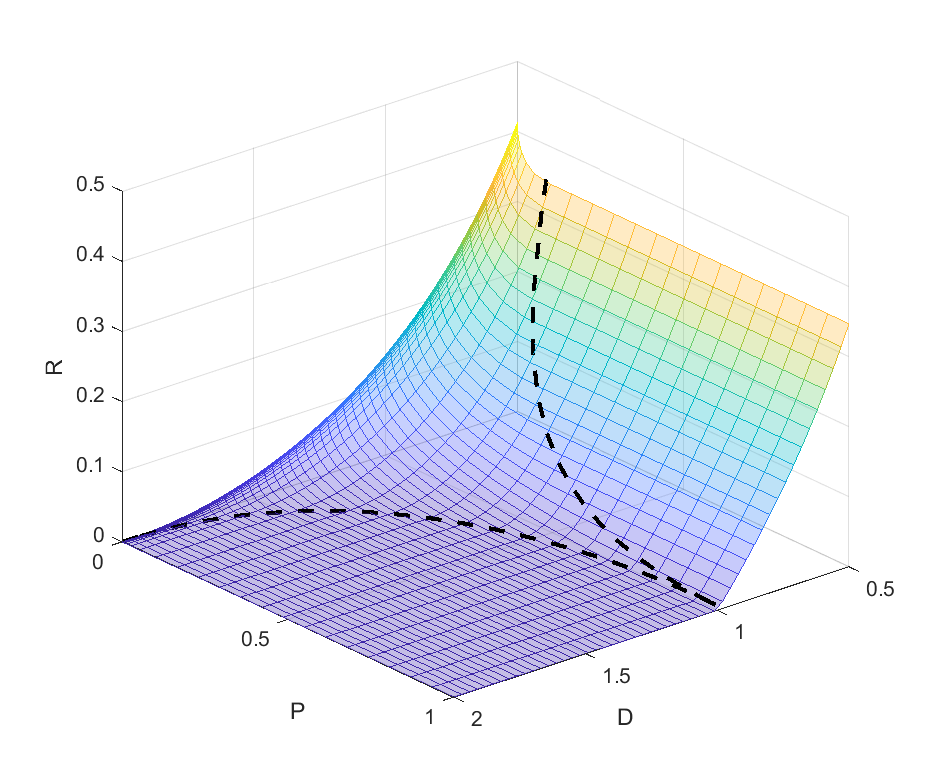}
        \caption{} \label{fig: uRDPW2}
    \end{subfigure}\\
\caption{$R(D,P)$ for a Gaussian source $X \sim \ND(0,1)$ source under  (a) direct $\KL$, (b)  reverse $\KL$, (c) $\GJS$, (d) $\HS$, and (e) $\WTS$  perception constraints.} 
\label{fig: UnivariateRDPFs}
\end{figure}

Let $X \sim \ND(0,1)$ be a scalar Gaussian source. Fig. \ref{fig: UnivariateRDPFs} shows the RDPF for $X$ under MSE distortion measure and, respectively, direct Kullback–Leibler divergence (\ref{fig: uRDPKLDirect}), reverse Kullback–Leibler divergence (\ref{fig: uRDPKLReverse}), Geometric Jensen-Shannon divergence (\ref{fig: uRDPGJS}), squared Hellinger distance (\ref{fig: uRDPHS}), and squared Wasserstein-2 distance (\ref{fig: uRDPW2}) perception measures.
 All the derived RDPFs share similarities in the structure of the operating regions on the $(D,P)$ plane. We distinguish three cases:
\begin{itemize}
    \item {\bf Case I}, where the RDPF function is identically similar to the associated RDF. In this regime, the perception constraint is not met with equality.
    \item {\bf Case II}, where, due to the distortion constraint not met with equality, the RDPF function is identically zero.
    \item {\bf Case III}, where both the distortion and perception constraints are met with equality.
\end{itemize}
In our derivations, we identify the operating regions $\{\text{\bf Case I} \} \cup \{\text{\bf Case II}\} = \mathcal{S}^c$ and $\{\text{\bf Case III}\} = \mathcal{S}$, giving us the closed-form solutions of their the operating regions.

{\subsection{Multivariate Gaussian Sources}}
This section is devoted to the analysis of the numerical results of Algorithm \ref{alg: AM}. All the numerical experiments in this section have been conducted considering a multivariate Gaussian source $X \sim \ND(0, \Cov[X])$ with $\Cov[X] = \diag([1, 3, 5])$ and setting the initialization point for Alg. \ref{alg: AM} to $(\vecD^{(0)},\vecP^{(0)}) = (0,0)$.

\subsubsection{RDPF Curves}
\begin{figure}[t]
    \centering
    \begin{subfigure}{\linewidth}
    	\centering \includegraphics[width=0.9\linewidth]{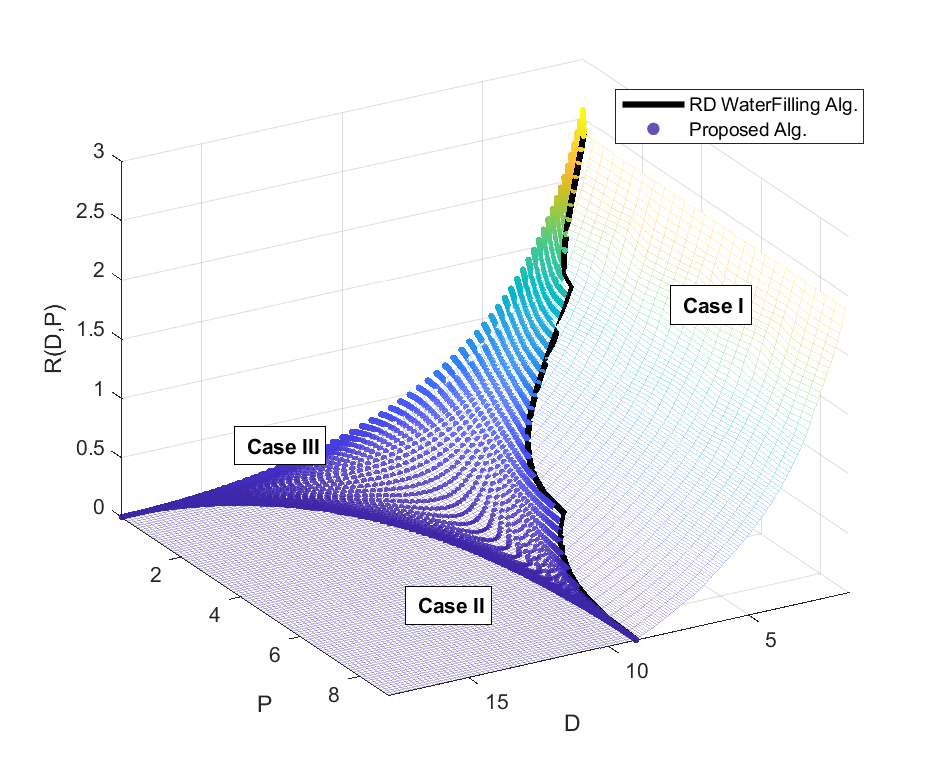}
    	\caption{} \label{fig: MRDPF_W2}
    \end{subfigure}\\
    \begin{subfigure}{0.49\linewidth}
        \centering \includegraphics[width=0.9\linewidth]{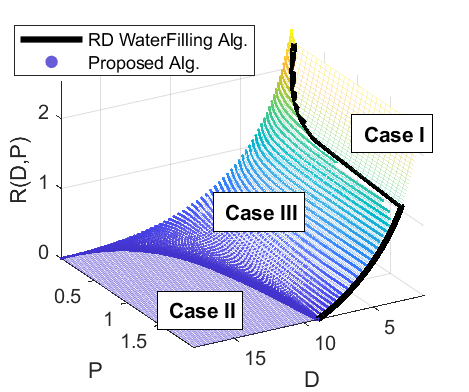}
        \caption{} \label{fig: MRDPF_HS}
    \end{subfigure}
    \begin{subfigure}{0.49\linewidth}
        \centering \includegraphics[width=0.9\linewidth]{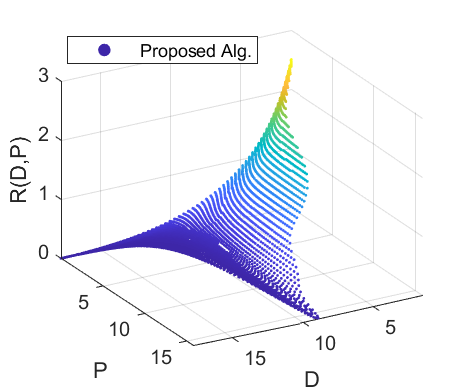}
        \caption{} \label{fig: MRDPF_KLd}
    \end{subfigure}\\
        \begin{subfigure}{0.49\linewidth}
        \centering \includegraphics[width=0.9\linewidth]{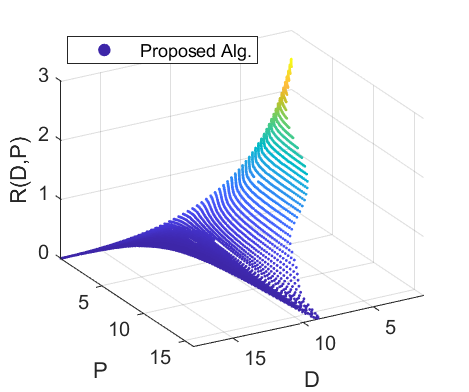}
        \caption{} \label{fig: MRDPF_KLr}
    \end{subfigure}
    \begin{subfigure}{0.49\linewidth}
        \centering \includegraphics[width=0.9\linewidth]{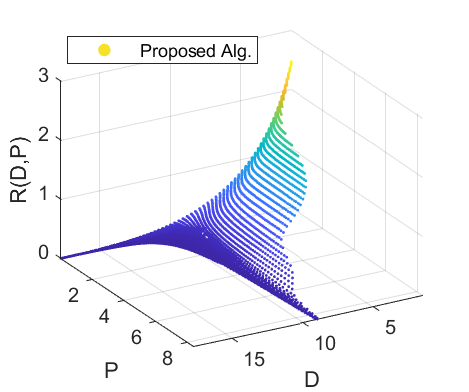}
        \caption{} \label{fig: MRDPF_GJS}
    \end{subfigure}\\
\caption{$R(D,P)$ for a Gaussian source $X \sim \ND(0,\Cov[X])$ with $\Cov[X] = \diag([1,3,5])$ under (a) $\WTS$ , (b) $\HS$, (c) direct $\KL$ , (d) reverse $\KL$, (d) $\HS$, and (e) $\GJS$  perception constraints.} 
\label{fig: MultivariateRDPFs}
\end{figure}

In Fig.~\ref{fig: MultivariateRDPFs}, we show the RDPF for the perception metrics introduced in Section \ref{sec: unvariate closed forms}. 
Focusing on Fig. \ref{fig: MRDPF_W2}, we compare the RDPF under squared Wasserstein-2 perception with the solution of the classical RD problem using water-filling algorithm (black line) where for the latter the perception measure has been computed a posteriori using the same divergence metric. The result confirms that for bounded divergence measure the RD solution can be obtained as extreme case of RDPF surface, retrieving the boundary between the regions of \textit{Case I} and \textit{Case III}. Moreover, the surface region of {\it Case I} can be retrieved by the rigid translation of the obtained boundary curve, since any $(D,P)$ point in this region defines an RDP problem where the perception constraint is not active, and thus equivalent to the classical RD problem. Similar remarks can be extended to the case of the RDPF under $\HS$ perception metric. In Figs.~\ref{fig: MRDPF_KLd}, \ref{fig: MRDPF_KLr}, and \ref{fig: MRDPF_GJS}, the same kind of comparison cannot be performed due to properties of the Kullback–Leibler and Geometric Jensen-Shannon divergences; depending on the distortion level $D$, the classical RD solution may induce the variance of one of the marginal distributions of the reconstructed source $\var[\hat{X}_i] \to 0$. In this condition, the absolute continuity between $\pdf{X}$ and $\pdf{\hat{X}}$ is not guaranteed, causing the measured perception level $P \to +\infty$. Since Algorithm \ref{alg: AM} presents identical behavior to the water-filling algorithm for $s_2 \to 0$ (equivalent to $P \to \infty$), for these cases the RDPFs have been computed bounding the maximum perception level $P$ imposing $s_2 \ge 10^{-3}$.

\subsubsection{Adaptive Water-Level}
\begin{figure}[htbp]
    \centering
    \includegraphics[width = \linewidth]{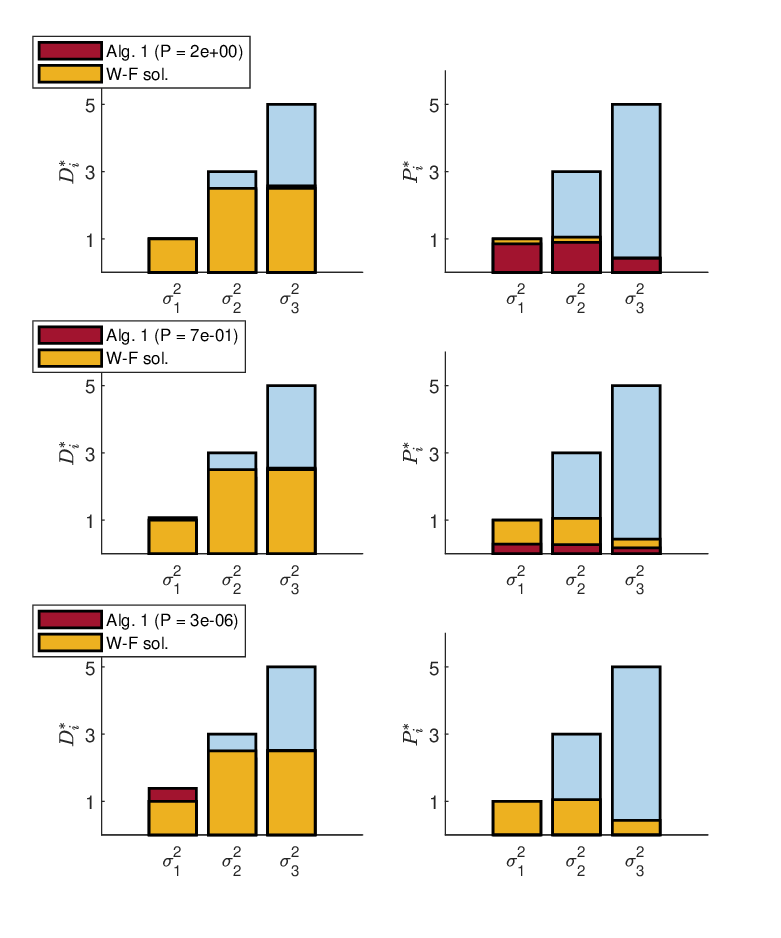}
    \caption{Comparison of the per-dimension distortion $D_i^*$ and perception $P_i^*$ measures for a fixed target distortion level $D = 6$ between the water-filling (WF) solution and Alg. \ref{alg: AM}. }
    \label{fig: comparison_per_dimension}
\end{figure}

In Fig. \ref{fig: comparison_per_dimension} we analyze the per-dimension levels of distortion $D_i^*$ and perception $P_i^*$ between the source $X$ and its reconstruction $\hat{X}$, comparing Algorithm \ref{alg: AM} with the classical RD water-filling solution. The comparison has been conducted with a target distortion level $D = 6$ and varying the target perception level $P$.

As anticipated in Remark \ref{remark: comparison with WF}, values of distortions $D^*_1 \ge \var[X_1]$ can be observed in the first dimension.

\appendices

{\section{Useful Results Obtained Via Tensorization}\label{sec:tensorization}}

{In this Appendix, we state and when needed prove certain useful results on tensorization of divergence functions and information measures.}

{First, we state a proposition which demonstrates how $\KL(\cdot||\cdot)$ can be tensorized. For a proof, see for instance \cite[Proposition 4]{fang:2020}.} 
\begin{proposition} \label{th:gmrdpf:KL_Commutation}
    Let $\Cov[X] \succ 0$ and $\Cov[\hX] \succ 0$ on $\mathbb{R}^{N \times N}$. {Moreover, let} $X \sim \ND(0,\Cov[X])$ and $\hX \sim \ND(0,\Cov[\hX])$. Then, $\KL(p_{X}||p_{\hat{X}})$ {can be bounded from below as follows}:
\begin{align*}
\KL(\pdf{X}||\pdf{\hX}) &\ge {\sum_{i=1}^N} \KL(\pdf{X_i}||\pdf{\hX_i})
\end{align*}
and the inequality holds with equality iff $\Cov[X]$ and $\Cov[\hX]$ are commuting matrices\footnote{The definition of commuting matrices can be found in \cite[Section 0.7.7]{horn:2012:matrix}.}.
\end{proposition}
{Similar to the case of KL divergence, the GJS divergence can also be tensorized. This is stated and proved next.}
\begin{proposition} \label{th:gmrdpf:GJS_Commutation}
    Let $\Cov[X] \succ 0$ and $\Cov[\hX] \succ 0$ on $\mathbb{R}^{N \times N}$. {Moreover, let} $X \sim \ND(0,\Cov[X])$ and $\hX \sim \ND(0,\Cov[\hX])$. Then, the GJS divergence $\GJS(\pdf{X}||\pdf{\hX})$ can be bounded {from below as follows}:
    \begin{align*}
        \GJS(\pdf{X}||\pdf{\hX}) &\ge {\sum_{i=1}^N} \GJS(\pdf{X_i}||\pdf{\hX_i})
    \end{align*}
{and the inequality holds with equality iff} $\Cov[X]$ and $\Cov[\hX]$ are commuting matrices.
\end{proposition}
\begin{IEEEproof}
{The tensorization of the $\GJS$ follows by observing that the geometric mean distribution $p_g$, as stated in Definition \ref{def: GJS divergence}, is itself Gaussian. Hence, if $\Cov[X]$ and $\Cov[\hX]$ commute, then $\Cov[g]$ necessarily commutes by pairs \cite[Section 0.7.7]{horn:2012:matrix}, thus allowing to directly apply \cite[Proposition 4]{fang:2020} to \eqref{eq:GJS_general} and obtain the result in question. This concludes the proof.}
\end{IEEEproof}
{Similarly, one can show that $\WTS$ can be tensorized. This is shown next.} 
\begin{proposition} \label{th:gmrdpf:W2_Commutation}
    Let $\Cov[X] \succ 0$ and $\Cov[\hX] \succ 0$ on $\mathbb{R}^{N \times N}$. {Moreover, let} $X \sim \ND(0,\Cov[X])$ and $\hX \sim \ND(0,\Cov[\hX])$. Then, $\WTS(\pdf{X},\pdf{\hX})$ can be bounded {from below as follows}:
    \begin{align}
      \WTS(\pdf{X},\pdf{\hX}) \ge {\sum_{i=1}^N} \WTS(\pdf{X_i},\pdf{\hX_i})\nonumber
    \end{align}
{and the inequality holds with equality iff} $\Cov[X]$ and $\Cov[\hX]$ are commuting matrices.
\end{proposition}
\begin{IEEEproof}
{For jointly Gaussian random variables,} the squared Wasserstein-2 distance $\WTS(\pdf{X},\pdf{\hX})$ has the following form \cite{gelbrich:1990:W2formula}:
\begin{align}
   { \WTS(\pdf{X},\pdf{\hX})} = \Tr \left[ \Cov[X] + \Cov[\hX] - 2 \Sqrt{\left( \Sqrt{\Cov[\hX]} \Cov[X]\Sqrt{\Cov[\hX]}\right)}\right]. \label{eq:Gaussian:W2_base}
\end{align}
{By performing eigenvalue decomposition on the covariance matrices $\Cov[X]$ and $\Cov[\hX]$ and writing them into their canonical form, i.e.,} $\Cov[X] = P\eigv{\Cov[X]}P^T$ and $\Cov[\hX] = Q\eigv{\Cov[\hX]}Q^T$, \eqref{eq:Gaussian:W2_base} can be expressed as follows:
\begin{align}
    \begin{split}
       { \WTS(\pdf{X},\pdf{\hX})} &= \Tr \left[\eigv{\Cov[X]}\right] + \Tr \left[ \eigv{\Cov[\hX]} \right] \\
        & \quad - 2 \Tr \left[ \Sqrt{\left( Q\Sqrt{\eigv{\Cov[\hX]}}Q^T P\eigv{\Cov[X]}P^TQ\Sqrt{\eigv{\Cov[\hX]}}Q^T\right)} \right].
    \end{split}\label{eq:W2_Tensor:W2_Gaussian}
\end{align} 
Our objective is the minimization of \eqref{eq:W2_Tensor:W2_Gaussian} while maintaining the set eigenvalues $\eigv{\Cov[X]}$ and $\eigv{\Cov[\hX]}$. Therefore, our optimization variables will be the orthonormal matrices \cite[Section 2.1]{horn:2012:matrix} $P$ and $Q$, whose columns are the set of eigenvectors of $\Cov[X]$ and $\Cov[\hat{X}]$, respectively. Since the first two terms of \eqref{eq:W2_Tensor:W2_Gaussian} do not depend of $P$ and $Q$, the minimization problem can be rewritten as a maximization problem as follows:
\begin{align}
    \max_{P,Q \in \mathcal{U}} \Tr \left[ \Sqrt{\left( Q\Sqrt{\eigv{\Cov[\hX]}}Q^T P\eigv{\Cov[X]}P^TQ\Sqrt{\eigv{\Cov[\hX]}}Q^T\right)} \right] \label{eq:Gaussian:W2Proof:tmp1}
\end{align}
where $\mathcal{U}$ is the set of orthonormal matrices on $\mathbb{R}^{N \times N}$. The optimization objective can be expressed as:
\begin{align*}
    &\Tr \left[ \Sqrt{\left( Q\Sqrt{\eigv{\Cov[\hX]}}Q^T P\eigv{\Cov[X]}P^TQ\Sqrt{\eigv{\Cov[\hX]}}Q^T\right)} \right]\\   &=||P\Sqrt{\eigv{\Cov[X]}}P^TQ\Sqrt{\eigv{\Cov[\hX]}}Q^T ||_1\\ 
    & \stackrel{(a)}{=}||\Sqrt{\eigv{\Cov[X]}}P^TQ\Sqrt{\eigv{\Cov[\hX]}}||_1
\end{align*}
where (a) is justified by the unitary invariance property of the norm. By defining $S \triangleq \Sqrt{\eigv{\Cov[X]}}P^TQ\Sqrt{\eigv{\Cov[\hX]}}$, we obtain
\begin{align*}
    ||S||_1 & \stackrel{(b)}{=} \Tr(|S|) \stackrel{(c)}{=} \sup_{U \in \mathcal{U}}\Tr(US)
\end{align*}
where (b) stems from the definition of Schatten-1 norm and by defining $|S| = \sqrt{S^TS}$ with $|S|$ being the unique positive semidefinite root of $S^TS$; (c) stems from the fact that using polar decomposition \cite[Theorem 7.3.1]{horn:2012:matrix}, there exists a unique matrix $U \in \mathcal{U}$ such that $|S| = US$ and the fact that $|S|$ maximizes $\Tr(US)$ over the set of unitary matrices $U \in \mathcal{U}$ \cite[Proposition 5]{bernal:19:maxTrace}.
{Consequentially, we can reformulate the optimization problem as follows:} 
\begin{align}
    \max_{P,Q \in \mathcal{U}} \sup_{U \in \mathcal{U}} \Tr\left(U\Sqrt{\eigv{\Cov[X]}}P^TQ\Sqrt{\eigv{\Cov[\hX]}}\right) \label{eq:Gaussian:W2Proof:tmp2}.
\end{align}
Now, let us define the function $f:\mathcal{U}\times\mathcal{U} \to \mathbb{R}$ as
\begin{align*}
    f(U,V) {\triangleq} \Tr(U\Sqrt{\eigv{\Cov[X]}}V^T\Sqrt{\eigv{\Cov[\hX]}}).\end{align*}
We can verify that $f$ {satisfies} the properties of an inner product on $\mathcal{U}$ \cite[Definition 5.1.3]{horn:2012:matrix}, therefore using Cauchy–Schwarz inequality \cite[Theorem 5.1.4]{horn:2012:matrix}, we obtain
\begin{align*}
\begin{split}
\sup_{U \in \mathcal{U}} \Tr\left(U\Sqrt{\eigv{\Cov[X]}}P^TQ\Sqrt{\eigv{\Cov[\hX]}}\right) &= \sup_{U \in \mathcal{U}} f(U,Q^TP) \\
& \le f(Q^TP, Q^TP)
\end{split}
\end{align*}
{Using the above inequality in \eqref{eq:Gaussian:W2Proof:tmp2} results into the following:}
\begin{align*}
    \eqref{eq:Gaussian:W2Proof:tmp2} & \stackrel{(d)}{\le} \max_{P,Q \in \mathcal{U}}  \Tr\left(Q^TP\Sqrt{\eigv{\Cov[X]}}P^TQ\Sqrt{\eigv{\Cov[\hX]}}\right) \\
    &= \max_{R = Q^TP \in \mathcal{U}} \Tr \left(R\Sqrt{\eigv{\Cov[X]}}R^T\Sqrt{\eigv{\Cov[\hX]}}\right)\\
    &= \max_{R = Q^TP \in \mathcal{U}} \langle \Sqrt{\eigv{\Cov[\hX]}}, R\Sqrt{\eigv{\Cov[X]}}R^T \rangle_F\\
    &\stackrel{(e)}{\le} \langle \Sqrt{\eigv{\Cov[\hX]}}, \Sqrt{\eigv{\Cov[X]}} \rangle_F
\end{align*}
where (d) and (e) hold with equality for {$Q^TP=I$}, implying {that the maximum of \eqref{eq:Gaussian:W2Proof:tmp1},  and consequentially the minimum of \eqref{eq:W2_Tensor:W2_Gaussian}, is attained by commuting matrices $\Cov[X]$ and $\Cov[\hX]$}. Therefore,
\begin{align*}
    { \WTS(\pdf{X},\pdf{\hX})} 
    &\ge \Tr \left[\eigv{\Cov[X]}\right] + \Tr \left[ \eigv{\Cov[\hX]} \right] - 2 \Tr \left[ \Sqrt{\eigv{\Cov[\hX]}}\Sqrt{\eigv{\Cov[X]}}\right]\\
    &= \sum_{i=1}^N \WTS(\pdf{X_i},\pdf{\hX_i}).
\end{align*}
{This concludes the proof.} 
\end{IEEEproof}

\begin{proposition} \label{th:gmrdpf:HS_Commutation}
    Let $\Cov[X] \succ 0$ and $\Cov[\hX] \succ 0$ on $\mathbb{R}^{N \times N}$. {Moreover, let} $X \sim \ND(0,\Cov[X])$ and $\hX \sim \ND(0,\Cov[\hX])$. Then, $\HS(\pdf{X},\pdf{\hX})$ can be bounded {from below as follows}:
    \begin{align}
      \HS(\pdf{X},\pdf{\hX}) \ge 2 \left(1 - \prod_{i=1}^N \left(1 - \frac{\HS(\pdf{X_i},\pdf{\hX_i})}{2} \right) \right)
    \end{align}
{and the inequality holds with equality iff} $\Cov[X]$ and $\Cov[\hX]$ are commuting matrices with increasing (or decreasing) eigenvalues order.
\end{proposition}
\begin{IEEEproof}
    For jointly Gaussian random variables, the squared $\HS(\pdf{X},\pdf{\hX})$ has the following form:
    \begin{align}
        \HS(\pdf{X},\pdf{\hX}) = 2\left( 1 - 2^{\frac{N}{2}}\frac{|\Cov[X]\Cov[\hX]|^{\frac{1}{4}}}{|\Cov[X] + \Cov[\hX]|^{\frac{1}{2}}}\right).
    \end{align}
    By performing eigenvalue decomposition on the covariance matrices $\Cov[X]$ and $\Cov[\hX]$ and writing them into their canonical form, i.e., $\Cov[X] = P\eigv{\Cov[X]}P^T$ and $\Cov[\hX] = Q\eigv{\Cov[\hX]}Q^T$, \eqref{eq:Gaussian:W2_base} can be expressed as follows:
    \begin{align}
        \HS(\pdf{X},\pdf{\hX}) &= 2\left( 1 - 2^{\frac{N}{2}}\frac{|P\eigv{\Cov[X]}P^TQ\eigv{\Cov[\hX]}Q^T|^{\frac{1}{4}}}{|P\eigv{\Cov[X]}P^T + Q\eigv{\Cov[\hX]}Q^T|^{\frac{1}{2}}}\right) \nonumber\\
        &=2\left( 1 - 2^{\frac{N}{2}}\frac{|\eigv{\Cov[X]}\eigv{\Cov[\hX]}|^{\frac{1}{4}}}{|\eigv{\Cov[X]} + R\eigv{\Cov[\hX]}R^T|^{\frac{1}{2}}}\right) \label{eq:HS:tensor:tmp1}
    \end{align}
with $R = P^TQ \in \mathcal{U}$, where $\mathcal{U}$ is the set of orthonormal matrices \cite[Section 2.1]{horn:2012:matrix}. Our objective is the minimization of \eqref{eq:HS:tensor:tmp1} while maintaining the set eigenvalues $\eigv{\Cov[X]}$ and $\eigv{\Cov[\hX]}$. Therefore, our optimization variables will be the orthonormal matrix $R$. Since only the denominator of the second term in \eqref{eq:HS:tensor:tmp1} depends on $R$, finding the minimizing $R$ is equivalent to
\begin{align*}
    \argmin_{R \in \mathcal{U}} |I + \iSqrt{\eigv{\Cov[X]}}R\eigv{\Cov[\hX]}R^T\iSqrt{\eigv{\Cov[\hX]}}|
\end{align*}
which is itself equivalent to
\begin{align}
    \argmin_{R \in \mathcal{U}} \log\left(|I + S^TS|\right) \label{eq:HS:tensor:tmp2}
\end{align}
where $S = \iSqrt{\eigv{\Cov[X]}}R\Sqrt{\eigv{\Cov[\hX]}}$. Now, expanding \eqref{eq:HS:tensor:tmp2} we obtain
\begin{align}
    \log\left(|I + S^TS|\right) & = \sum_{i=1}^N \log(\lambda[I + S^TS]) \nonumber\\
    &\stackrel{(a)}= \sum_{i=1}^N \log( 1 + \lambda_i[S^TS]) \nonumber\\
    &\stackrel{(b)}= \sum_{i=1}^N \log( 1 + \sv^2_{S,i})
\end{align}
where $(a)$ follows from the fact that for any diagonalizable matrix $M$ the eigenvalues $\lambda_{I+M} = 1 + \lambda_{M}$ and $(b)$ follows from the definition of singular values of a matrix. In addition, we can also verify the following majorization inequality: 
\begin{align}
    \begin{split}
        &\log\left( \sv \left[ \iSqrt{\eigv{\Cov[X]}}R\Sqrt{\eigv{\Cov[\hX]}} \right] \right)\\
        & \stackrel{(c)}{\succ} \log\left( \sv^{\downarrow} \left[ \iSqrt{\eigv{\Cov[X]}}R\right] \right) + \log \left( \lambda ^{\uparrow} \left[\Cov[X]^{\tfrac{1}{2}} \right] \right) \\
        & \stackrel{(d)}{\succ} \log\left( \lambda ^{\downarrow} \left[\Cov[X]^{-\tfrac{1}{2}} \right] \right) + \log \left( \sv^{\uparrow} \left[ R \right] \right) + \log \left(  \lambda ^{\uparrow} \left[\Cov[X]^{\tfrac{1}{2}} \right]\right) \\
        &\stackrel{(e)}{=} \log\left( \frac{1}{\lambda^{\uparrow} \left[ \Cov[X]^{\tfrac{1}{2}}\right]} \right) + \log \left( \lambda ^{\uparrow} \left[\Cov[\hX]^{\tfrac{1}{2}} \right]\right)
    \end{split} \label{eq:HS:tensor:eq2}
\end{align}
where (c) and (d) hold from Lidskii's Theorem \cite[Theorem 3.4.6]{bhatia:2013} and hold with equality for commuting matrices $\eigv{\Cov[X]}$, $\eigv{\Cov[\hX]}$ and $R$ with the correct singular value ordering, while (e) follows from $R$ being orthonormal, hence $\sv_{R,i} = 1 ~\forall i\in1:N$. Now, let $f(\cdot)$ to be the function defined as $f(x)\triangleq \log( 1 + x^2)$. Although $f$ is not convex, we can readily show that the composite function $t \xrightarrow{} f(e^t)$ is convex and increasing on $\mathbb{R}$. Thus, we can invoke Weyl's majorization theorem, which states:
\begin{align}
\log(x) \succ_w \log(y) \implies f(x) \succ_w f(y). \label{eq:HS:tensor:eq3}
\end{align}
Therefore, we obtain
\begin{align*}
   \eqref{eq:HS:tensor:tmp2}  =  \sum_{i = 1}^N f \left(\sv_{i} \left[\iSqrt{\eigv{\Cov[X]}}R\Sqrt{\eigv{\Cov[\hX]}}\right] \right) \stackrel{\eqref{eq:HS:tensor:eq2} + \eqref{eq:HS:tensor:eq3}}{\ge} \sum_{i = 1}^N f \left( \frac{\stveigvi[\hX] ^{\uparrow}}{\stveigvi[X]^{\uparrow}} \right) 
\end{align*}
where the lower bound is achievable iff the matrix $R$ induces the proper ordering between $\eigv{\Cov[X]}$ and $\eigv{\Cov[\hX]}$, including the case where $\Cov[X]$ and $\Cov[\hX]$ are commuting matrices with eigenvalues in increasing (or decreasing) order. Therefore, assuming that both $\Cov[X]$ and $\Cov[\hX]$ have the same eigenvalues order, \eqref{eq:HS:tensor:tmp1} is lower bounded by
\begin{align*}
    \HS(\pdf{X},\pdf{\hX}) &\ge 2\left( 1 - \prod_{i=1}^N \sqrt{\frac{2\stveigvi[X]\stveigvi[\hX]}{\eigvi{\Cov[X]} + \eigvi{\Cov[\hX]}}}\right)\\
    & = 2 \left(1 - \prod_{i=1}^N \left(1 - \frac{\HS(\pdf{X_i},\pdf{\hX_i})}{2} \right) \right)
\end{align*}
where $\HS(\pdf{X_i},\pdf{\hX_i})$ is the squared Hellinger distance between the marginals $X_i \sim \ND(0, \eigvi{\Cov[X]})$ and $\hX_i \sim \ND(0, \eigvi{\Cov[\hX]})$. This concludes the proof.

\end{IEEEproof}

\par {Next, we derive a tensorization result for mutual information of jointly Gaussian random vectors.} 

\begin{lemma} \label{th:gmrdpf:MI_Commutation}
  {Let $(X, \hX)$ be jointly Gaussian random vectors on $\mathbb{R}^N$, such that $X \sim \ND(0,\Cov[X])$ with $\Cov[X] \succ 0$ and $\hX = AX + W$, with $A\in\mathbb{R}^{N \times N}$ being invertible and diagonalizable, and $W \sim \ND\left(0, \Cov[W] \right)$ with $\Cov[W] \succ 0$. Then, the mutual information $I(X,\hX)$ can be bounded from below as follows:  
    \begin{align}
        I(X,\hX) = \log \left( \frac{|\Cov[\hX]|}{|\Cov[\hX|X]|} \right) \ge \sum_{i=1}^N \log \left( 1 + \frac{\lambda_{A,i}^2 \lambda_{\Cov[X],i}}{\lambda_{\Cov[W],i}} \right) \label{gmrdpf:th_rate:lowebound}
    \end{align}
and the inequality holds with equality iff $A$ is a symmetric matrix and the matrices $\Cov[X]$, $\Cov[W]$ and $A$ commute by pairs.}
\end{lemma}
\begin{IEEEproof}
Note that for jointly Gaussian random vectors $(X, \hX)$, mutual information between $X$ and $\hat{X}$ can be written as follows:
\begin{align}
    I(X,\hX)&=\log \left( \frac{|\Cov[\hX]|}{|\Cov[\hX|X]|} \right)  \nonumber\\ 
    & = \log \left( \frac{|A\Cov[X]A^T + \Cov[W]|}{|\Cov[W]|} \right) \nonumber\\
    & = \log \left( |I + \Cov[W]^{-\tfrac{1}{2}}A\Cov[X]A^T\Cov[W]^{-\tfrac{1}{2}}| \right) \nonumber\\
    & = \log \left( \left|I + \left( \Cov[W]^{-\tfrac{1}{2}}A\Cov[X]^{\tfrac{1}{2}} \right) \left(\Cov[W]^{-\tfrac{1}{2}}A\Cov[X]^{\tfrac{1}{2}} \right)^T \right| \right) \nonumber\\
    & = \log \left( \left|I + \left(\Cov[W]^{-\tfrac{1}{2}}A\Cov[X]^{\tfrac{1}{2}} \right)^T \left( \Cov[W]^{-\tfrac{1}{2}}A\Cov[X]^{\tfrac{1}{2}} \right)  \right| \right) \nonumber\\
    &\stackrel{(a)}= \sum_{i = 1}^N \log \left( 1 + \lambda_{i} \left[\left(\Cov[W]^{-\tfrac{1}{2}}A\Cov[X]^{\tfrac{1}{2}} \right)^T \left(\Cov[W]^{-\tfrac{1}{2}}A\Cov[X]^{\tfrac{1}{2}} \right) \right] \right) \nonumber\\
    &\stackrel{(b)} = \sum_{i = 1}^N \log \left( 1 + \left(\sv_{i} \left[\Cov[W]^{-\tfrac{1}{2}}A\Cov[X]^{\tfrac{1}{2}} \right] \right)^2 \right) \label{gmrdpf:th_rate:eq1}
\end{align}
{where $(a)$ follows from the fact that for any diagonalizable matrix $M$ the eigenvalues $\lambda_{I+M} = 1 + \lambda_{M}$ and $(b)$ follows from the definition of singular values of a matrix}. In addition, we can also verify the following majorization inequality: 
\begin{align}
    \begin{split}
        &\log\left( \sv \left[ \Cov[W]^{-\tfrac{1}{2}}A\Cov[X]^{\tfrac{1}{2}} \right] \right)\\
        & \stackrel{(c)}{\succ} \log\left( \sv^{\downarrow} \left[ \Cov[W]^{-\tfrac{1}{2}}A \right] \right) + \log \left( \sv^{\uparrow} \left[\Cov[X]^{\tfrac{1}{2}} \right] \right) \\
        & \stackrel{(d)}{\succ} \log\left( \sv^{\downarrow} \left[ \Cov[W]^{-\tfrac{1}{2}}\right] \right) + \log \left( \sv^{\uparrow} \left[ A \right] \right) + \log \left( \sv^{\uparrow} \left[\Cov[X]^{\tfrac{1}{2}} \right] \right) \\
        & \stackrel{(e)}{\succ_w} \log\left( \lambda^{\downarrow} \left[ \Cov[W]^{-\tfrac{1}{2}}\right] \right) + \log \left( |\lambda^{\uparrow} \left[ A \right]| \right) + \log \left( \lambda ^{\uparrow} \left[\Cov[X]^{\tfrac{1}{2}} \right] \right)\\
        &= \log\left( \frac{1}{\lambda^{\uparrow} \left[ \Cov[W]^{\tfrac{1}{2}}\right]} \right) + \log \left( |\lambda^{\uparrow}_A| \right) + \log \left( \lambda ^{\uparrow} \left[\Cov[X]^{\tfrac{1}{2}} \right]\right)
    \end{split} \label{gmrdpf:th_rate:eq3}
\end{align}
where (c) and (d) hold from Lidskii's theorem \cite[Theorem 3.4.6]{bhatia:2013} and hold with equality for commuting matrices $\Cov[X]$, $\Cov[W]$ and $A$ with the correct singular value ordering, while (e) follows from Weyl's majorization theorem \cite[Theorem 2.3.6]{bhatia:2013} and holds with equality {if and only if} $\Cov[X]$, $A$, and $\Cov[W]$ are a symmetric matrices. \\
Now, let $f(\cdot)$ to be the function defined as $f(x)\triangleq \log( 1 + x^2)$. Although $f$ is not convex, we can readily show that the composite function $t \xrightarrow{} f(e^t)$ is convex and increasing on $\mathbb{R}$. Thus, we can invoke Weyl's majorization theorem, which states:
\begin{align}
\log(x) \succ_w \log(y) \implies f(x) \succ_w f(y). \label{gmrdpf:th_rate:eq2}
\end{align}
Therefore, we obtain:
\begin{align*}
   \eqref{gmrdpf:th_rate:eq1}  &=  \sum_{i = 1}^N f \left(\sv_{i} \left[\Cov[W]^{-\tfrac{1}{2}}A\Cov[X]^{\tfrac{1}{2}}\right] \right)\\
   &\stackrel{\eqref{gmrdpf:th_rate:eq2} + \eqref{gmrdpf:th_rate:eq3}}{\ge} \sum_{i = 1}^N f \left( \frac{|\eigvi{A}^{\uparrow}| \cdot \stveigvi[X] ^{\uparrow}}{\stveigvi[W]^{\uparrow}} \right) = \eqref{gmrdpf:th_rate:lowebound}. 
\end{align*}
This concludes the proof.
\end{IEEEproof}

{We conclude this section by showing that the MSE admits a lower bound via tensorization for jointly Gaussian random vectors.}
\begin{lemma}\label{th:gmrdpf:MSE_Commutation}
{Let $(X, \hX)$ be jointly Gaussian random vectors on $\mathbb{R}^N$  such that $X\sim\ND(0,\Cov[X])$ and $\hX = AX + W$, with $A\in\mathbb{R}^{N \times N}$ being invertible and diagonalizable, and $W \sim \ND\left(0, \Cov[W] \right)$. Moreover, let $\Cov[X]$ and $\Cov[W]$ be  positive definite matrices. Then, the MSE can be bounded from below as follows:
    \begin{align*}
        \E{||X - \hX||^2} \ge \sum_{i=1}^N \E{(X_i - \hX_i)^2}
    \end{align*}
whereas the inequality holds with equality if $A$ is symmetric and ($A, \Cov[W], \Cov[X])$ commute by pairs.}
\end{lemma}
\begin{IEEEproof}
{Note that we can expand the MSE as follows:
\begin{align}
\E{||X - \hX||^2} &= \Tr \left( \Cov[X] + \Cov[\hX] - 2A\Cov[X] \right)\nonumber\\
                    &\stackrel{(a)}= \Tr \left[ (I-A)\Cov[X](I-A)^T + \Cov[W] \right] \nonumber\\
                    &= \Tr \left[ \left( (I-A)\Sqrt{\Cov[X]} \right)^T\left((I-A)\Sqrt{\Cov[X]} \right) + \Cov[W] \right]\nonumber\\
                    &\stackrel{(b)} = ||(I-A)\Sqrt{\Cov[X]}||_F^2 + \Tr[\Cov[W]]\label{proof:mse_eq.1}
\end{align}
{where $(a)$ follows from the substitution $\Cov[\hX] = A\Cov[X]A^T + \Cov[W]$ and $(b)$ follows from the definition of Frobenius norm $||\cdot||_F$}.
Moreover, from \eqref{proof:mse_eq.1} we can derive
\begin{align*}
    ||(I-A)\Sqrt{\Cov[X]}||_F^2 
    & = \sum_{i = 1}^N \sv_i^2\left[(I-A)\Sqrt{\Cov[X]}\right]\\
    & \stackrel{(c)}{\ge} \sum_{i = 1}^N  \left( \sv_i^{\downarrow}\left[I-A\right]\right)^2 \lambda_i^{\uparrow}(\Cov[X])     \\
    & \stackrel{(d)}{\ge} \sum_{i = 1}^N  \left( |\lambda_i^{\downarrow}\left[I-A\right]|\right)^2  \lambda_i^{\uparrow}(\Cov[X])    \\       
    & \ge \sum_{i = 1}^N  \left( |1 - \lambda_i^{\uparrow}\left[A\right]| \right)^2 \lambda_i^{\uparrow}(\Cov[X])
\end{align*}
where (c) follows from Lidskii's theorem \cite[Theorem 3.4.6]{bhatia:2013} and hold with equality for commuting matrices $(A,\Cov[X])$ with the correct singular value ordering, while (d) follows from Weyl's majorization theorem \cite[Theorem 2.3.6]{bhatia:2013} and holds with equality {if and only if} $A$ is a symmetric matrix. This concludes the proof.}
\end{IEEEproof}

{\section{Proof of Theorem \ref{th:uRDP:KL}} \label{proof:uRDP:KL}}

{First note that once we substitute} the functional form of the direct {$\KL(\pdf{X}||\pdf{\hX})$} in \eqref{opt: uRDPGaussian}, {we obtain the following convex optimization problem}:
\begin{align}
     R(D,P) &=  \min_{a\in\mathbb{R},\var[W]\ge{0}} \tfrac{1}{2} \log(1 + a^2 \frac{\var[X]}{\var[W]}). \label{eq:directKL:obj}\\
    \textrm{s.t.}   & \quad (1-a)^2\var[X] + \var[W] \le D \label{eq:directKL:MSE} \\
                    \tfrac{1}{2} \Biggr[\log &\left( \frac{a^2\var[X] + \var[W]}{\var[X]} \right) + \frac{\var[X]}{a^2\var[X] + \var[W]} - 1 \Biggl] \le P.\label{eq:directKL:KL}
\end{align}
{To solve the specific optimization problem, we invoke Karush-Kuhn-Tucker (KKT) conditions \cite{boyd:2004}}. Let $s_1\ge{0}$ and $s_2\ge{0}$ be the Lagrangian multipliers associated with the distortion and perception constraint, respectively. The Lagrangian function $L(a,\var[W], s_1, s_2)$ associated with \eqref{eq:directKL:obj}-\eqref{eq:directKL:KL} can give the unconstrained optimization problem as follows:
\begin{align*}
    &L(a,\var[W], s_1, s_2) = \tfrac{1}{2} \log(1 + a^2 \frac{\var[X]}{\var[W]})\\
    & \quad + s_1( (1-a)^2\var[X] + \var[W] - D) \\
    & \quad + \frac{s_2}{2}\left\{\Biggr[\log \left( \frac{a^2\var[X] + \var[W]}{\var[X]} \right) + \frac{\var[X]}{a^2\var[X] + \var[W]} - 1\Biggl] - P \right\} .
\end{align*}
The {\it stationarity conditions} and {\it complementary slackness} are as follows:
\begin{align}
    &\nabla_{(a,\var[W])} L(a,\var[W], s_1, s_2) = 0  \label{eq:directKL:stationarity}\\
    &s_1( (1-a)^2\var[X] + \var[W] - D) = 0 \label{eq:directKL:SlackD}\\
    &\frac{s_2}{2}\left\{\Biggr[\log \left( \frac{a^2\var[X] + \var[W]}{\var[X]} \right) + \frac{\var[X]}{a^2\var[X] + \var[W]} - 1\Biggl] - P \right\} = 0 \label{eq:directKL:SlackP}.
\end{align}
Now, we obtain the complete closed-form solution for the convex programming problem in \eqref{eq:directKL:obj}-\eqref{eq:directKL:KL} by breaking it into three distinct cases.\\
{\textbf{Case I:} Suppose that $s_1>0$ and $s_2=0$, namely, only \eqref{eq:directKL:MSE} is active in the optimization problem. Then, the problem is equivalent to the classical RD (see, e.g., \cite{berger:1971}), which has known optimal realization $\hat{X}=aX+W$ with design variables $(a,\var[W])$ given as follows:
 \begin{align}
     a &= \left(1 - \tfrac{D}{\var[X]} \right),~\var[W] = D - (1-a)^2\var[X]. \label{proof:KLD:design_var:CaseI}
 \end{align}
\textbf{Case II:} Let $s_1=0$ and $s_2>0$, namely, only \eqref{eq:directKL:KL} is active in the optimization problem. Then, from \eqref{eq:directKL:SlackP}, we obtain:
\begin{align}
    \log\left( \frac{\var[\hX]}{\var[X]} \right) + \frac{\var[X]}{\var[\hX]} &= 2P + 1\nonumber\\
    \Longrightarrow-\frac{\var[X]}{\var[\hX]} \cdot e^{-\frac{\var[X]}{\var[\hX]}} &= -e^{-(2P + 1)}\nonumber\\
    \Longrightarrow\frac{\var[X]}{\var[\hX]} &= -\lamb[]\left(-e^{-(2P + 1)}\right) \label{eq:derivation:KLD:Wfunction}
\end{align}
where \eqref{eq:derivation:KLD:Wfunction} is obtained through the application of the Lambert $W$ function \cite{corless:1996:lambert}}. From \eqref{eq:directKL:stationarity} and \eqref{eq:directKL:SlackP}, we can derive the optimal $(a,\var[W])$, resulting in: 
\begin{align}
    a = 0,~~\var[W] = -\frac{\var[X]}{\lamb[]\left(-e^{-(2P + 1)}\right)}\label{proof:design_var:2}
\end{align}
which, once substituted in \eqref{eq:directKL:obj}, shows that the inactive distortion constraint results in $R(D,P) = 0$.\\ 
\noindent{\textbf{Case III:} Suppose that $s_1>0$ and $s_2>0$, namely, both \eqref{eq:directKL:MSE} and \eqref{eq:directKL:KL} are active. Then, from \eqref{eq:directKL:SlackD} we have that $a^2\var[X] + \var[W] = D + (2a-1)\var[X]$ and by substituting in \eqref{eq:directKL:SlackP} we obtain:}
 \begin{align*}
     \log \left( \tfrac{D + (2a-1)\var[X]}{\var[X]} \right) + \tfrac{\var[X]}{D + (2a-1)\var[X]} &= 2P + 1\\
     \Longrightarrow-\tfrac{\var[X]}{D + (2a-1)\var[X]} \cdot e^{-\tfrac{\var[X]}{D + (2a-1)\var[X]}} &= -e^{-(2P + 1)}\\
     \Longrightarrow\tfrac{D + (2a-1)\var[X]}{\var[X]} &= -\lamb[](-e^{-(2P + 1)})
 \end{align*}
{that, upon further simplifications, results into the following choice of the design variables $(a,\var[W])$:}
\begin{align}
     a = \tfrac{1}{2}\left(1 - \tfrac{D}{\var[X]} - \tfrac{1}{\lamb[](-e^{-(2P + 1)})}  \right),~     \var[W] = D - (1-a)^2\var[X],\label{proof:design_var:3}
 \end{align}
 which once substituted in \eqref{eq:directKL:obj} {give \eqref{eq:RDP_GaussianKL}}.
{It remains to determine domain boundaries for each of the  three cases and to identify which branch of the Lambert $W$ function makes the derived $R(D,P)$ compatible with the properties listed in Remark \ref{remark:1}.}
\par {Note that the boundary between \textbf{Case I} and \textbf{Case III} can be identified by considering only \textbf{Case I} and by substituting its solutions in \eqref{eq:directKL:KL}, which yields}
 \begin{align}
     -\tfrac{\var[X]}{\var[X] - D} \cdot e^{-\tfrac{\var[X]}{\var[X] - D}} &< -e^{-(2P + 1)}. \label{eq:directKL:boundary13temp}
 \end{align}
 Applying the Lambert $W$ function to inequality \eqref{eq:directKL:boundary13temp} requires a decision between the distinct branches $\lamb[0]$ and $\lamb[-1]$, since the two have opposite monotonic behaviors ({that is,} monotonically increasing and monotonically decreasing, {respectively}). However, invoking the continuity of the RDPF, we select the branch of the Lambert $W$ function resulting in $R(D,P)$ being continuous on border. Of the two branches, only $\lamb[-1]$ respects this continuity constraint. Therefore, \eqref{eq:directKL:boundary13temp} becomes
 \begin{align}
    \frac{\var[X]}{\var[X] - D} < - \lamb[-1]\left(-e^{-(2P + 1)}\right).\label{eq:KL:boundary13}
\end{align}
{Similarly, one can identify the boundary between \textbf{Case II} and \textbf{Case III} by considering solely \textbf{Case II} and substituting its solution in \eqref{eq:directKL:MSE}, which results into the following strict inequality:}
\begin{align}
    \frac{\var[X]}{\var[X] - D} > \lamb[-1]\left(-e^{-(2P + 1)}\right).\label{eq:KL:boundary23}
\end{align}
From inequalities \eqref{eq:KL:boundary13} and \eqref{eq:KL:boundary23} we can identify the joint region of \textbf{Case I} and \textbf{Case II} on the $(D,P)$ plane, denoted by the set $\mathcal{S}^c = \{ (D,P): \eqref{eq:KL:boundary13} \cup \eqref{eq:KL:boundary23} \}$. Subsequently, the region of \textbf{Case III} is identified as $\mathcal{S} = \left(\mathcal{S}^c\right)^c$, resulting in \eqref{RDP_DKL:CaseIIIRegion}. {This concludes the proof.}

{\section{Proof of Lemma \ref{th:uRDP:W2}} \label{proof:uRDP:W2}}
We only sketch the proof as it uses the same logical structure of the proof of Theorem \ref{th:uRDP:KL}. The $\WTS$ distance between two scalar Gaussian random variables $X \sim \ND(0, \var[X])$ and $\hX \sim \ND(0, \var[\hX])$ takes form
\begin{align}
    \WTS(\pdf{X},\pdf{\hX}) = \left(\stv[X] - \stv[\hX] \right)^2.\label{eq:W2:uW2}
\end{align}
By substituting \eqref{eq:W2:uW2} in \eqref{opt: uRDPGaussian}, we obtain the following convex optimization problem
\begin{align}
     R(D,P)=\min_{a,\var[W]}& \tfrac{1}{2} \log(1 + a^2 \tfrac{\var[X]}{\var[W]}) \label{eq:W2:obj}\\
    \textrm{s.t.}   & \quad (1 - a^2)\var[X] + \var[W] \le D \label{eq:W2:MSE} \\
                    & \quad \eqref{eq:W2:uW2} \le P. \label{eq:W2:Perception} 
\end{align}
Let $s_1\ge{0}$ and $s_2\ge{0}$ be the Lagrangian multipliers associated with the distortion and perception constraint, respectively. Similarly to Theorem \ref{th:uRDP:KL}, we will obtain the closed-form solution of the optimization problem in \eqref{eq:W2:obj}-\eqref{eq:W2:Perception} by leveraging the KKT conditions \cite{boyd:2004}, considering specific operating regions of the problem and identifying the sets where these solutions apply.\\
\textbf{Case I:} Suppose $s_1 > 0$ and $s_2 = 0$, namely, only \eqref{eq:W2:MSE} is active in the optimization problem. Then, the problem becomes equivalent to the classical $RD$, therefore, as in Theorem \ref{th:uRDP:KL}, the optimal design variables $(a,\var[W])$ are equal to \eqref{proof:KLD:design_var:CaseI}.\\
\textbf{Case II:} Let $s_1 = 0$ and $s_2 > 0$, namely only \eqref{eq:W2:Perception} holds with equality. As in Theorem \ref{th:uRDP:KL}, the inactive distortion constraint \eqref{eq:W2:MSE} implies that $R(D,P) = 0$ on this operating region. Moreover, we can design the design variables $(a,\var[W])$ as:
\begin{align}
    a = 0, ~\var[W] = (\stv[X] - \sqrt{P})^2. \label{W2:sol:case2}
\end{align}
\textbf{Case III:} Suppose $s_1 > 0$ and $s_2 > 0$, namely both \eqref{eq:W2:MSE} and \eqref{eq:W2:Perception} are active and holding with equality. We can observe from \eqref{eq:W2:MSE} that:
\begin{align}
    \var[W] &= D - (1 - a)^2\var[X] \label{eq:W2:sigmaW} \\
    \var[\hX] &= a^2\var[X] + \var[W] = D + (2a-1)\var[X] \label{eq:W2:temp1}
\end{align}
and from \eqref{eq:W2:Perception} we obtain:
 \begin{align}
        \left| \stv[X] - \sqrt{a^2\var[X] + \var[W]} \right| = \sqrt{P}  \label{eq:W2:temp3}
 \end{align}
 Note that \eqref{eq:W2:temp3} behaves as follows:
 \begin{itemize}
     \item if $\stv[X] \ge \sqrt{a^2\var[X] + \var[W]} $:
     \begin{align}
         a = \tfrac{1}{2} \frac{\var[X] + (\stv[X] - \sqrt{P})^2 -D}{\var[X]} \label{eq:W2:a_set1}
     \end{align}
     \item if $\stv[X] < \sqrt{a^2\var[X] + \var[W]} $:   
     \begin{align}
        a = \tfrac{1}{2} \frac{\var[X] + (\stv[X] + \sqrt{P})^2 -D}{\var[X]} \label{eq:W2:a_set2}
     \end{align}
 \end{itemize}
Between \eqref{eq:W2:a_set1} and \eqref{eq:W2:a_set2}, the former holds the minimum once substituted in \eqref{eq:W2:obj} with \eqref{eq:W2:sigmaW}. Therefore, the optimal design variables $(a,\var[W]) = \left( \eqref{eq:W2:a_set1}, \eqref{eq:W2:sigmaW} \right)$. 
\par Regarding the boundaries between the operating regions, similarly to Theorem \ref{th:uRDP:KL} the boundary between \textbf{Cases I-III} can be obtained by considering \textbf{Case I} and substituting its solution \eqref{proof:KLD:design_var:CaseI} in \eqref{eq:W2:Perception}, while the boundary between \textbf{Cases II-III} can be obtained by substituting \eqref{W2:sol:case2} in \eqref{eq:W2:MSE}. This  concludes the proof.

{\section{Proof of Theorem \ref{th:uRDP:GJS}} \label{proof:uRDP:GJS}}

The proof follows the same approach as the proof of Theorem \ref{th:uRDP:KL}, therefore we highlight only the differences between the two while maintaining the same logical structure.\\
The $\GJS$ divergence \eqref{eq:GJS_general} between two scalar Gaussian random variables $X \sim \ND(0, \var[X])$ and $\hX \sim \ND(0, \var[\hX])$ is defined as
\begin{align}
    \GJS(\pdf{X}||\pdf{\hX}) \triangleq \tfrac{1}{4}\left[ -\log \left( \tfrac{1}{4}\tfrac{(\var[X] + \var[\hX])^2}{\var[X]\var[\hX]}\right) + \tfrac{1}{2}\tfrac{(\var[X] + \var[\hX])^2}{\var[X]\var[\hX]} -2\right]. \label{eq:GHS:GJS}
\end{align}
Once \eqref{eq:GHS:GJS} is substituted in  \eqref{opt: uRDPGaussian}, we obtain the following convex optimization problem:
\begin{align}
     R(D,P)=\min_{a,\var[W]}& \tfrac{1}{2} \log(1 + a^2 \tfrac{\var[X]}{\var[W]}) \label{eq:GJS:obj}\\
    \textrm{s.t.}   & \quad (1 - a^2)\var[X] + \var[W] \le D \label{eq:GJS:MSE} \\
                    & \quad \eqref{eq:GHS:GJS} \le P. \label{eq:GJS:Perception} 
\end{align}
  Similarly to Theorem \ref{th:uRDP:KL}, we obtain the closed-form solution of the optimization problem in \eqref{eq:GJS:obj}-\eqref{eq:GJS:Perception} by leveraging the KKT conditions, considering specific operating regions of the problem and identifying the sets where these solutions apply. Let $s_1\ge{0}$ and $s_2\ge{0}$ be the Lagrangian multipliers associated with the distortion and perception constraint, respectively.\\
 \textbf{Case I:} Let $s_1 > 0$ and $s_2 = 0$, namely, only \eqref{eq:GJS:MSE} is active in the optimization problem. Then, the problem becomes equivalent to the classical $RD$, therefore, as in Theorem \ref{th:uRDP:KL}, the optimal design variables $(a,\var[W])$ are equal to \eqref{proof:KLD:design_var:CaseI}.\\
\textbf{Case II:} Suppose $s_1 = 0$ and $s_2 > 0$, namely only \eqref{eq:GJS:Perception} holds with equality. As in Theorem \ref{th:uRDP:KL}, the inactive distortion constraint \eqref{eq:GJS:MSE} implies that $R(D,P) = 0$ on this operating region. Moreover, we can design the design variables $(a,\var[W])$ as:
\begin{align}
    a = 0, ~\var[W] = -\frac{\var[X]}{1 + 2\var[X]\lamb[-1](-2e^{-(4P+2)})}. \label{GJS:sol:case2}
\end{align}
\textbf{Case III:} Let $s_1 > 0$ and $s_2 > 0$, meaning that both \eqref{eq:GJS:MSE} and \eqref{eq:GJS:Perception} are active and hold with equality. We can observe from \eqref{eq:GJS:MSE} that
\begin{align}
    \var[\hX] = a^2\var[X] + \var[W] = D + (2a-1)\var[X] \label{eq:GJS:temp1}
\end{align}
and from \eqref{eq:GJS:Perception} we obtain
 \begin{align*}
     \log \left( \tfrac{(\var[X] + \var[\hX])^2}{\var[\hX]\var[X]} \right) -\tfrac{(\var[X] + \var[\hX])^2}{2\var[\hX]\var[X]} &=  -(4P +2 -\log(4))\\
     -\tfrac{(\var[X] + \var[\hX])^2}{2\var[\hX]\var[X]} \cdot e^{-\tfrac{(\var[X] + \var[\hX])^2}{2\var[\hX]\var[X]}} &= -2 \cdot e^{-(4P + 2)}\\
     \tfrac{(\var[X] + \var[\hX])^2}{\var[\hX]\var[X]} &= -2\lamb[](-2 \cdot e^{-(4P + 2))}).
 \end{align*}
As in Theorem \ref{th:uRDP:KL}, it is necessary to understand which branch of the Lambert $W$ respects the continuity property of RDPF. For the sake of simplicity, we will directly use the negative branch $\lamb[-1]$ which can be later proved to be the correct one.
Defining $g(P) = -2\lamb[-1](-2 \cdot e^{-(4P + 2))})$, the equation expands two a second degree polynomial in $\var[\hX]$:
\begin{align*}
    (\var[\hX])^2 + \var[\hX]\var[X](2-g(P)) + (\var[X])^2 = 0.
\end{align*}
The solutions of the polynomial are both positive for $P\in[0,+\infty)$ and we select the one minimizing $\var[\hX]$, thus obtaining:
\begin{align*}
    \var[\hX] = \tfrac{1}{2}\var[X]\left[ -(2-g(P)) - \sqrt{g(P)(g(P) -4)}\right].
\end{align*}
Enforcing \eqref{eq:GJS:temp1} and solving for $a$ and $\var[W]$ results in
\begin{align}
     a &= \tfrac{1}{2}\left[1 - \tfrac{D}{\var[X]} - \left((2-g(P)) + \sqrt{g(P)(g(P) -4)}\right)  \right]\\
     \var[W] &= D - (1-a)^2\var[X]
 \end{align}
 which once substituted in \eqref{eq:GJS:obj} give  \eqref{eq:RDP_GaussianGJS}.
\par Regarding the boundaries between the operating regions, similarly to Theorem \ref{th:uRDP:KL}, the boundary between \textbf{Cases I-III} can be obtained by considering \textbf{Case I} and substituting its solution \eqref{proof:KLD:design_var:CaseI} in \eqref{eq:GJS:Perception}, while the boundary between \textbf{Cases II-III} can be obtained by substituting \eqref{GJS:sol:case2} in \eqref{eq:GJS:MSE}. This  concludes the proof.

{\section{Proof of Theorem \ref{th:uRDP:HS}} \label{proof:uRDP:HS}}

The proof follows the same approach as the proof of Theorem \ref{th:uRDP:KL}, therefore we highlight only the differences between the two while maintaining the same logical structure.\\
The $\HS$ distance \eqref{eq:HS_general} between two scalar Gaussian random variables $X \sim \ND(0, \var[X])$ and $\hX \sim \ND(0, \var[\hX])$ is defined as:
\begin{align}
    \HS(\pdf{X},\pdf{\hX}) \triangleq 1 - \sqrt{\frac{\stv[X]\stv[\hX]}{\var[X] + \var[\hX]}}. \label{eq:HS:HS}
\end{align}
Once \eqref{eq:HS:HS} is substituted in  \eqref{opt: uRDPGaussian}, we obtain the following convex optimization problem:
\begin{align}
     R(D,P)=\min_{a,\var[W]}& \tfrac{1}{2} \log(1 + a^2 \tfrac{\var[X]}{\var[W]}) \label{eq:HS:obj}\\
    \textrm{s.t.}   & \quad (1 - a^2)\var[X] + \var[W] \le D \label{eq:HS:MSE} \\
                    & \quad \eqref{eq:HS:HS} \le P. \label{eq:HS:Perception} 
\end{align}
  Similarly to Theorem \ref{th:uRDP:KL}, we obtain the closed-form solution of the optimization problem in \eqref{eq:HS:obj}-\eqref{eq:HS:Perception} by leveraging the KKT conditions, considering specific operating regions of the problem and identifying the sets where these solutions apply. Let $s_1\ge{0}$ and $s_2\ge{0}$ be the Lagrangian multipliers associated with the distortion and perception constraint, respectively.\\
 \textbf{Case I:} Let $s_1 > 0$ and $s_2 = 0$, namely, only \eqref{eq:HS:MSE} is active in the optimization problem. Then, the problem becomes equivalent to the classical $RD$, therefore, as in Theorem \ref{th:uRDP:KL}, the optimal design variables $(a,\var[W])$ are equal to \eqref{proof:KLD:design_var:CaseI}.\\
\textbf{Case II:} Suppose $s_1 = 0$ and $s_2 > 0$, namely only \eqref{eq:HS:Perception} holds with equality. As in Theorem \ref{th:uRDP:KL}, the inactive distortion constraint \eqref{eq:HS:MSE} implies that $R(D,P) = 0$ on this operating region. Moreover, we can design the design variables $(a,\var[W])$ as:
\begin{align}
    a = 0, ~\var[W] = \var[X] \left(1 - 2g(P) \right). \label{HS:sol:case2}
\end{align}
\textbf{Case III:} Let $s_1 > 0$ and $s_2 > 0$, meaning that both \eqref{eq:HS:MSE} and \eqref{eq:HS:Perception} are active and hold with equality. We can observe from \eqref{eq:HS:MSE} that
\begin{align}
    \var[\hX] = a^2\var[X] + \var[W] = D + (2a-1)\var[X] \label{eq:HS:temp1}
\end{align}
and from \eqref{eq:HS:Perception} we obtain
 \begin{align}
    \var[\hX] = \var[X] \left( \frac{1 \pm \sqrt{1 - (1-P)^4}}{(1-P)^2 }\right)^2. \label{eq:HS:temp2}
 \end{align}
Now, solving the equation $\eqref{eq:HS:temp1} - \eqref{eq:HS:temp2} = 0$ for $a$ holds two feasible solutions, of which we select the one minimizing \eqref{eq:HS:obj}. Hence, the optimal design variables $(a,\var[W])$ are the following:
\begin{align}
     a  =  g(P) - \frac{D}{2\var[X]}, ~\var[W] = D - (1-a)^2\var[X].
 \end{align}
\par Regarding the boundaries between the operating regions, similarly to Theorem \ref{th:uRDP:KL} the boundary between \textbf{Cases I-III} can be obtained by considering \textbf{Case I} and substituting its solution \eqref{proof:KLD:design_var:CaseI} in \eqref{eq:HS:Perception}, while the boundary between \textbf{Cases II-III} can be obtained by substituting \eqref{HS:sol:case2} in \eqref{eq:HS:MSE}. This  concludes the proof.

{\section{Proof of Theorem \ref{th:gmrdpf:RDPF_AM}} \label{proof:gmrdpf:RDPF_AM}}
{To prove this result, we use some prior results of Grippo and Sciandrone in \cite{grippo:2000} in the study of the 2-block non-linear Gauss-Seidel method under convex constraints.}
\par {Let $f(\vecD,\vecP) = \sum_{i=1}^N R_i(D_i,P_i)$ and let the set $\mathcal{L}_{(D,P)}$ be the set of level curves defined as follows:
  \begin{align}
    \mathcal{L} = \left\{ (\vecD,\vecP) \in \mathcal{C}_{(D,P)}: f(\vecD,\vecP) \le f(\vecD^{(0)},\vecP^{(0)}) \right\}\nonumber
\end{align}
where $\mathcal{C}_{(D,P)} = \{(\vecD,\vecP) \in \mathbb{R}^{2N}: \sum_{i=1}^N g(D_i) \le D \land \sum_{i = 1}^N h(P_i) \le P\}$. Since the pair $(D,P)$ is assumed to be finite, we remark that $\mathcal{C}_{(D,P)}$ is a compact set, and, as $f(\cdot)$ is a non-negative, continuous function in $\mathcal{C}_{(D,P)}$, the curve level set $\mathcal{L} = f^{-1}([0,f(\vecD^{(0)},\vecP^{(0)})]) \cap \mathcal{C}_{(D,P)}$ is also a compact set. This result together with the convexity of $f(\cdot)$ satisfy the assumption of \cite[Proposition 6]{grippo:2000}, which guarantee that every limit point $(\vecD^{*},\vecP^{*})$ of the sequence $\{(\vecD^{(n)},\vecP^{(n)}):~n=1,2,\ldots\}$ is a global minimizer of  \eqref{opt:gmrdpf:full}. This concludes the proof.}

\section{Proof of Theorem \ref{th:conv:convrate}} \label{proof:conv:convrate}
 The Lagrangian function $L(\vecD,\vecP,s_1,s_2)$ associated with \eqref{opt:gmrdpf:full} is an unconstrained convex function defined as follows:
\begin{align*}
    L(\vecD,\vecP) &= \sum_{i=1}^N R_i(D_i,P_i) + s_1(\sum_{i=1}^N D_i - D)\\
    &\quad + s_2(\sum_{i=1}^N P_i - P).
\end{align*}
Since $R_i(\cdot,\cdot)$ is continuous with Lipschitz continuous gradients in both arguments, there exist $K \in \mathbb{R}^+$ such that the Lagrangian $L(\vecD,\vecP)$ is continuous and its gradients $\nabla_\vecD L(\cdot, \vecP)$ and $\nabla_\vecP f(\vecD, \cdot)$ are $K$-Lipschitz continuous. Therefore, due to convexity the following holds $\forall ~ \vecD_1, \vecD_2, \vecP_1, \vecP_2 \in \mathbb{R}^N$:
\begin{align}
     L(\vecD_2, \vecP_1) &\le L(\vecD_1, \vecP_1) + \nabla_\vecD f(\vecD_1, \vecP_1)^T(\vecD_2 - \vecD_1) \nonumber \\
    & \quad + \frac{K}{2}||\vecD_2 - \vecD_1 ||^2_2. \label{diseq:conv:convex_D}
\end{align}
Alg. \ref{alg: AM} induces the following alternating minimization scheme $ \vecP^{(n-1)} \to \vecD^{(n)} \to \vecP^{(n)} $ obtained by the alternating solutions of the optimization problems \eqref{opt:gmrdpf:partial_optD} and \eqref{opt:gmrdpf:partial_optP}. Considering iteration $n$, for fixed $\vecP^{(n)}$ we define
\begin{align*}
    \vecD^{(n+1)} &= \argmin_\vecD ~ \eqref{opt:gmrdpf:partial_optD}\\
    \tilde{\vecD} &= \vecD^{(n)} - \nabla_\vecD L  (\vecD^{(n)}, \vecP^{(n)})
\end{align*}
for which it holds by construction
\begin{align}
    L(\tilde{\vecD}, \vecP^{(n)}) \ge L(\vecD^{(n+1)}, \vecP^{(n)}). \label{diseq:conv:gradient_descent}
\end{align}
From \eqref{diseq:conv:convex_D} and \eqref{diseq:conv:gradient_descent} we can derive:
\begin{align}
    \begin{split}
        &L(\vecD^{(n)}, \vecP^{(n)}) - L(\vecD^{(n+1)}, \vecP^{(n)}) \\
        & \ge L(\vecD^{(n)}, \vecP^{(n)}) - L(\tilde{\vecD}, \vecP^{(n)})\\
        & \ge -\nabla_\vecD f(\vecD_1, \vecP_1)^T(\tilde{\vecD} - \vecD^{(n)}) - \frac{K}{2}||\tilde{\vecD} - \vecD^{(n)} ||^2_2\\
        & \ge \frac{1}{2K}||\nabla_\vecD L  (\vecD^{(n)}, \vecP^{(n)})||_2^2. 
    \end{split} \label{diseq:conv:main_D}
\end{align}
 Conversely, we can derive a similar bound to \eqref{diseq:conv:main_D} considering iteration $n+1$ with fixed $\vecD^{(n+1)}$, obtaining:
 \begin{align}
    \begin{split}
        L(\vecD^{(n+1)}, \vecP^{(n)}) &\ge L(\vecD^{(n+1)}, \vecP^{(n+1)}) \\
        &+ \frac{1}{2K}||\nabla_\vecP L  (\vecD^{(n+1)}, \vecP^{(n)})||_2^2. 
    \end{split} \label{diseq:conv:main_P}
\end{align}
Combining \eqref{diseq:conv:main_D} and \eqref{diseq:conv:main_P}, we obtain
\begin{align*}
    &L(\vecD^{(n)}, \vecP^{(n)}) - L(\vecD^{(n+1)}, \vecP^{(n+1)}) \ge \frac{1}{2K} \omega^2_n\\
    &\omega^n = \left(||\nabla_\vecD L  (\vecD^{(n)}, \vecP^{(n)})||_2^2 + ||\nabla_\vecP L  (\vecD^{(n+1)}, \vecP^{(n)})||_2^2\right).
\end{align*}
Summing over $T$ iterations, we obtain
\begin{align*}
    \begin{split}
        &L(\vecD^{(0)}, \vecP^{(0)}) - L(\vecD^{(T+1)}, \vecP^{(T+1)}) \\
        &\ge \frac{1}{2K} \sum_{n=0}^T \omega^2_n  \ge \frac{T}{2K} \min_{n \in 0:T} \omega^2_n.
    \end{split}
\end{align*}
Since $L(\vecD,\vecP)$ is lower bounded by $R(D,P)$ with $D = \sum_{i = 0}^N D_i$ and $P = \sum_{i = 0}^N P_i$, we obtain
\begin{align*}
    \begin{split}
        &\frac{\sqrt{2KC}}{\sqrt{T}} \ge \min_{n \in 0:T} \omega_n
    \end{split}
\end{align*}
where $C = L(\vecD^{(0)}, \vecP^{(0)}) - R(D,P) \ge 0$ is a constant that depends only on the initial point $(\vecD^{(0)}, \vecP^{(0)})$ and the operating point $(D,P)$ is identified by the associated Lagrangian multipliers $(s_1,s_2)$. This concludes the proof.

\section{Proof of Theorem \ref{th:gmrdpf:AM_subD}} \label{proof:gmrdpf:AM_subD}

    The Lagrangian $L(s_1, \vecD)$ associated with \eqref{opt:gmrdpf:partial_optD} has the following unconstrained form:
    \begin{align*}
        L(s_1, \vecD) = \sum_{i = 1}^N R_i(D_i,P_i) + s_1 \left(\sum_{i = 1}^N D_i - D \right).
    \end{align*}
    Applying the \textit{stationarity condition} from the KKT conditions we obtain
    \begin{align}
        \frac{\partial L(s_1, D_i)}{\partial D_i}\Bigg|_{D_i^*} = \frac{\partial R_i(D_i,P_i)}{\partial D_i}\Bigg|_{D_i^*} + s_1 = 0 \label{th:gmrdpf:subop1:KKTConditions}
    \end{align}
    with
    \begin{align*}
        &\frac{\partial R_i(D_i,P_i)}{\partial D_i} = \\
        &\begin{cases}
            -\frac{1}{2D_i} &\qquad (D_i, P_i) \in \mathcal{S}^c\\
            - \frac{D_i - P_i - 2\eigvi{\Cov[X]} + 2\sqrt{P_i\eigvi{\Cov[X]}}}{(D_i - P_i)(D_i - (2\sqrt{\eigvi{\Cov[X]}} - \sqrt{P_i})^2} &\qquad (D_i, P_i) \in \mathcal{S}
        \end{cases}
    \end{align*}
    where  $\mathcal{S}$ is defined in Theorem \ref{th:uRDP:W2}. We can investigate the optimal solution of \eqref{th:gmrdpf:subop1:KKTConditions} in the two sets $\mathcal{S}$ and $\mathcal{S}^c$ as follows:
    \begin{itemize}
        \item for $(D_i^*, P_i) \in \mathcal{S}$ we have that
        \begin{align*}
            -\frac{1}{2D^*_i} + s_1 = 0 \implies D_i^* = \frac{1}{2s_1}
        \end{align*}
        \item for $(D_i^*, P_i) \in \mathcal{S}^c$ we have that
        \begin{align*}
            - \frac{D_i - P_i - 2\eigvi{\Cov[X]} + 2\sqrt{P_i\eigvi{\Cov[X]}}}{(D_i - P_i)(D_i - (2\sqrt{\eigvi{\Cov[X]}} - \sqrt{P_i})^2} + s_1 = 0
        \end{align*}
        which admits the following two solutions:
        \begin{align*}
            &\begin{split}
                D_{i,1}^*   &= P_i + 2 \stv[X](\stv[X] - \sqrt{P_i}) \\
                        & \quad + \left(\frac{1}{2s_1} - \sqrt{4\var[X](\stv[X] - \sqrt{P_i})^2 + \tfrac{1}{4s_1^2}} \right)
            \end{split}\\
            &\begin{split}
                D_{i,2}^*   &=  P_i + 2 \stv[X](\stv[X] - \sqrt{P_i}) \\
                        & \quad + \left(\frac{1}{2s_1} + \sqrt{4\var[X](\stv[X] - \sqrt{P_i})^2 + \tfrac{1}{4s_1^2}} \right).
            \end{split}
        \end{align*}
        We select the solution $D_{i}^* = D_{i,1}^*$ since $D^*_{i,1} < D^*_{i,2}$, minimizing $D = \sum_{i = 1}^N D_i^*$, and this guarantees continuity of the function $D^*_i (s_1)$ on $\mathcal{S} \cup \mathcal{S}^c = \mathbb{R}^{2,+}$. This completes the proof.
    \end{itemize}
    
\section{Proof Theorem \ref{th:gmrdpf:AM_subP}} \label{proof:gmrdpf:AM_subP}

The Lagrangian $L(s_2, \vecP)$ associated with \eqref{opt:gmrdpf:partial_optP} has form:
\begin{align*}
    L(s_2, \vecP) = \sum_{i = 1}^N R_i(D_i,P_i) + s_2\left(\sum_{i = 1}^N P_i - P\right).
\end{align*}
We can characterize the set of optimal solutions $\{\vecP^*\}$ imposing the KKT conditions:
\begin{align*}
    \frac{\partial L(s_2, P_i)}{\partial P_i}\Bigg|_{P_i^*} = \frac{\partial R_i(D_i,P_i)}{\partial P_i}\Bigg|_{P_i^*} + s_2 = 0. \label{th:gmrdpf:subop1:KKTConditions}
\end{align*}
We notice that set $\{\vecP^*\}$ and the set of the zeros of $T(\cdot)$ coincide. This concludes the proof.

\section{Proof Corollary \ref{lemma:PropertiesTi}} \label{proof:PropertiesTi}

Since $R_i(D_i,P_i)$ is convex and differentiable in both arguments, $\tfrac{\partial R_i(D_i,P_i)}{\partial P_i}$ is necessarily continuous and non-decreasing on $\mathbb{R}_0^+$. From the definition of the set $\mathcal{S}$, we can characterize $b_i \ge 0$ such that $\mathcal{S} = [0, b_i]$ and $\mathcal{S}^c = (b_i, +\infty)$. Due to continuity, $\tfrac{\partial R_i(D_i,b_i)}{\partial P_i} = 0$, thus implying $T_i(b_i) \ge 0$. On the other hand, $\lim_{P_i \to 0} \tfrac{\partial R_i(D_i,b_i)}{\partial P_i} = - \infty$ implies the existence of $a_i \in \mathcal{S}$ such that $T_i(a_i) \le 0$. Hence, since on the extreme of the set $[a_i, b_i] \subseteq \mathcal{S}$ the continuous function $T_i$ has opposite signs, by the Intermediate Value Theorem \cite[Theorem 1.11]{burden:2015} there exists at least one root in the set. This concludes the proof.

\bibliographystyle{IEEEtran}
\bibliography{strings, biblio}

\end{document}

%% file: mathDeclarations.tex
\DeclareMathOperator{\id}{id}
\DeclareMathOperator{\diag}{diag}

\DeclareMathOperator{\IID}{i.i.d.}

\DeclareMathOperator*{\argmin}{arg\,min}
\DeclareMathOperator{\Tr}{Tr}
\DeclareMathOperator{\KL}{D_{KL}}
\DeclareMathOperator{\GJS}{D_{GJS}}
\DeclareMathOperator{\WTS}{W_2^2}
\DeclareMathOperator{\HS}{H^2}
\DeclareMathOperator{\MSE}{MSE}
\DeclareMathOperator{\ND}{\mathcal{N}}

\newcommand{\eigv}[1]{ \lambda_{#1} }
\newcommand{\sv}{s}
\newcommand{\eigvi}[1]{{\lambda_{#1,i}}}
\newcommand{\eigviSq}[1]{{\lambda^2_{#1,i}}}
\newcommand{\mean}[1][]{ \ifthenelse{\isempty{#1}}{\mu}{\mu_{#1}}}
\newcommand{\var}[1][]{\ifthenelse{\isempty{#1}}{\sigma^2} {\sigma^2_{#1}}}
\newcommand{\Cov}[1][]{\ifthenelse{\isempty{#1}}{\Sigma} {\Sigma_{#1}}}
\newcommand{\Sqrt}[1]{#1^{\frac{1}{2}}}
\newcommand{\iSqrt}[1]{#1^{-\frac{1}{2}}}
\newcommand{\stv}[1][]{\ifthenelse{\isempty{#1}}{\sigma}{\sigma_{#1}}}
\newcommand{\stveigvi}[1][]{\ifthenelse{\isempty{#1}}{\sqrt{\eigvi{\Cov[X]}}}{\sqrt{\eigvi{\Cov[#1]}}}}
\newcommand{\lamb}[1][]{\ifthenelse{\isempty{#1}}{\mathcal{W}} {\mathcal{W}_{#1}}}
\newcommand{\hX}{\hat{X}}

\newcommand{\pdf}[1]{ p_{#1} }
\newcommand{\cpdf}[2]{P_{#1|#2}}
\newcommand{\vecD}{\mathbf{D}}
\newcommand{\vecP}{\mathbf{P}}

\newcommand{\E}[2][]{ 
    \mathbb{E}_{#1} \left[ #2 \right]
}

\newcommand{\rem}[1]{\textcolor{orange}{#1}}

\newtheorem{theorem}{\bfseries Theorem}
\newtheorem{lemma}{\bfseries Lemma}
\newtheorem{definition}{\bfseries Definition}
\newtheorem{corollary}{\bfseries Corollary}

\newtheorem{remark}{\bfseries Remark}

\newtheorem{proposition}{\bfseries Proposition}
\newtheorem{example}{\bfseries Example}

\newtheorem*{assumptions*}{\bfseries Assumptions}

\newtheorem{problem}{\bfseries Problem}